\documentclass[sn-nature]{sn-jnl}

\usepackage{lineno}
\usepackage{graphicx}%
\usepackage{multirow}%
\usepackage{amsmath,amssymb,amsfonts}%
\usepackage{amsthm}%
\usepackage{mathrsfs}%
\usepackage[title]{appendix}%
\usepackage{xcolor}%
\usepackage{textcomp}%
\usepackage{manyfoot}%
\usepackage{booktabs}%
\usepackage{algorithm}%
\usepackage{algorithmicx}%
\usepackage{algpseudocode}%
\usepackage{listings}%
\DeclareUnicodeCharacter{2061}{}

\usepackage{graphicx}%
\usepackage{multirow}%
\usepackage{amsmath,amssymb,amsfonts}%
\usepackage{amsthm}%
\usepackage{physics}
\usepackage{mathrsfs}%
\usepackage[title]{appendix}%
\usepackage{xcolor}%
\usepackage{textcomp}%
\usepackage{manyfoot}%
\usepackage{booktabs}%
\usepackage{algorithm}%
\usepackage{algorithmicx}%
\usepackage{algpseudocode}%
\usepackage{listings}%
\DeclareUnicodeCharacter{2061}{}

\begin{document}

\title[Resource-efficient photonic quantum computation with high-dimensional cluster states]{Resource-efficient photonic quantum computation with high-dimensional cluster states}

\author*[1]{\fnm{Ohad} \sur{Lib}}\email{ohad.lib@mail.huji.ac.il}

\author[1]{\fnm{Yaron} \sur{Bromberg}}

\affil*[1]{\orgdiv{Racah Institute of Physics}, \orgname{The Hebrew University of Jerusalem}, \orgaddress{ \city{Jerusalem}, \postcode{91904}, \country{Israel}}}

\abstract{Quantum computers can revolutionize science and technology, but their realization remains challenging across all platforms. A promising route to scalability is photonic measurement-based quantum computation, where single-qubit measurements on large cluster states, together with feedforward, enable fault-tolerant quantum computation. However, generating large cluster states at high rates is notoriously difficult, as detection probabilities drop exponentially with the number of photons comprising the state. We tackle this challenge by encoding multiple qubits on each photon through high-dimensional spatial encoding, generating cluster states with over nine qubits at a rate of 100Hz. Additionally, we demonstrate that high-dimensional encoding substantially reduces the computation duration by enabling instantaneous feedforward between qubits encoded in the same photon. Our findings pave the way for resource-efficient measurement-based quantum computation using high-dimensional entanglement.}

\maketitle
From simulating quantum systems\cite{daley2022practical} to factoring large numbers\cite{shor1994algorithms}, quantum computers promise substantial speedups over their classical counterparts. Their realization, however, has proven to be extremely challenging across all physical platforms. In matter-based systems, decoherence and noise limit the achievable circuit depth\cite{preskill2018quantum}, whereas, in photonic systems, weak photon-photon interaction renders two-qubit gates difficult\cite{knill2001scheme,rudolph2017optimistic}. A potential remedy for the lack of efficient two-photon gates is the measurement-based quantum computing (MBQC) paradigm\cite{raussendorf2001one}. In MBQC, a large initial entangled state called a cluster state\cite{briegel2001persistent} is first generated, and universal quantum computation is then performed through simple single-qubit measurements and classical feedforward without the need for two-qubit gates\cite{raussendorf2003measurement}.

However, while MBQC circumvents the need to perform photonic two-qubit gates during the computation, it does so by shifting the difficulty to the generation of large cluster states at high rates. Indeed, despite significant theoretical\cite{browne2005resource,toth2005detecting,kok2007linear,bartolucci2021creation}, experimental\cite{walther2005experimental,lu2007experimental,yao2012experimental,thomas2022efficient}, and technological\cite{natarajan2012superconducting} advances since the first demonstration of entanglement in the 1970s\cite{freedman1972experimental,aspect1982experimental}, state-of-the-art realizations of photonic cluster states are still limited to roughly ten qubits at sub-Hz detection rates\cite{yao2012experimental,thomas2022efficient}. This poor scalability is mainly due to the exponential decrease in the generation and detection probabilities when the number of photons comprising the state is increased, which stems from both fundamental difficulties in entangling non-interacting photons\cite{yao2012experimental}, and technical issues, such as finite collection and detection efficiencies\cite{thomas2022efficient,cogan2023deterministic}. The exponential scaling with the number of photons sets a fundamental trade-off in current experiments between the number of qubits in the cluster state and its detection rate, limiting the scale of MBQC demonstrations.

A promising approach for generating large cluster states without increasing the number of photons is to encode the quantum information using hyper- or high-dimensional entanglement\cite{fickler2012quantum,kues2017chip,martin2017quantifying,valencia2020high}. In this case, the large Hilbert space associated with each photon can encode multiple qubits or high-dimensional quantum bits (qudits), thus reducing the number of required photons\cite{erhard2020advances}. In the context of MBQC, early demonstrations based on polarization-spatial hyper-entanglement have shown increased generation rates of small cluster states\cite{vallone2007realization,chen2007experimental,ceccarelli2009experimental} and reduced feedforward overhead\cite{vallone2008active}. More recently, the generation of noise-resilient $d=3$ qudit cluster states using temporal-spectral encoding\cite{reimer2019high} and the encoding of error-protected qubits in integrated photonic chips\cite{vigliar2021error,zhang2023encoding} were demonstrated.

Increasing the Hilbert space dimension of each photon brings new challenges to the generation, manipulation, and certification of cluster states, due to the large number of optical modes that need to be precisely controlled. In addition, since the measurement of a single photon now collapses multiple qubits simultaneously, performing complete measurement-based gates that require feedforward on a single photon becomes nontrivial and could naively only be done probabilistically. As a result, a clear advantage of high-dimensional MBQC over current state-of-the-art two-dimensional systems has not been observed so far.

Here, we experimentally generate large cluster states with up to $9.28$ qubits, that simultaneously exhibit a high detection rate of 100Hz. We overcome the size-rate trade-off of binary encoding by entangling multiple qudits within the Hilbert space of each photon, thus enlarging the state without sacrificing its generation efficiency. We achieve this by utilizing the spatial degree of freedom of entangled photons, which we precisely control through ten bounces off a spatial light modulator. This demonstrates the advantage of high-dimensional spatial encoding for photonic quantum computation, joining recent works in quantum communication\cite{mirhosseini2015high}, quantum networks\cite{leedumrongwatthanakun2020programmable,goel2022inverse,network2023science}, and entanglement certification\cite{ecker2019overcoming,srivastav2022quick}. In addition, we show that the simultaneous collapse of multiple qubits upon the measurement of their encoding photon is, in fact, beneficial for implementing feedforward between them. We perform such feedforward instantaneously using tailored linear optic circuits and demonstrate a potential four-fold reduction in the duration of single-qubit rotations in MBQC. Our approach therefore reduces the required experimental resources in terms of both the number of photons and active feedforward hardware, highlighting the potential of high-dimensional encoding for resource-efficient MBQC on different photonic platforms\cite{yao2012experimental,thomas2022efficient,istrati2020sequential,cogan2023deterministic}.

\section*{Results}
\subsection*{Spatially encoded cluster states}

\begin{figure}[H]
\centering
\includegraphics[width=0.9\textwidth]{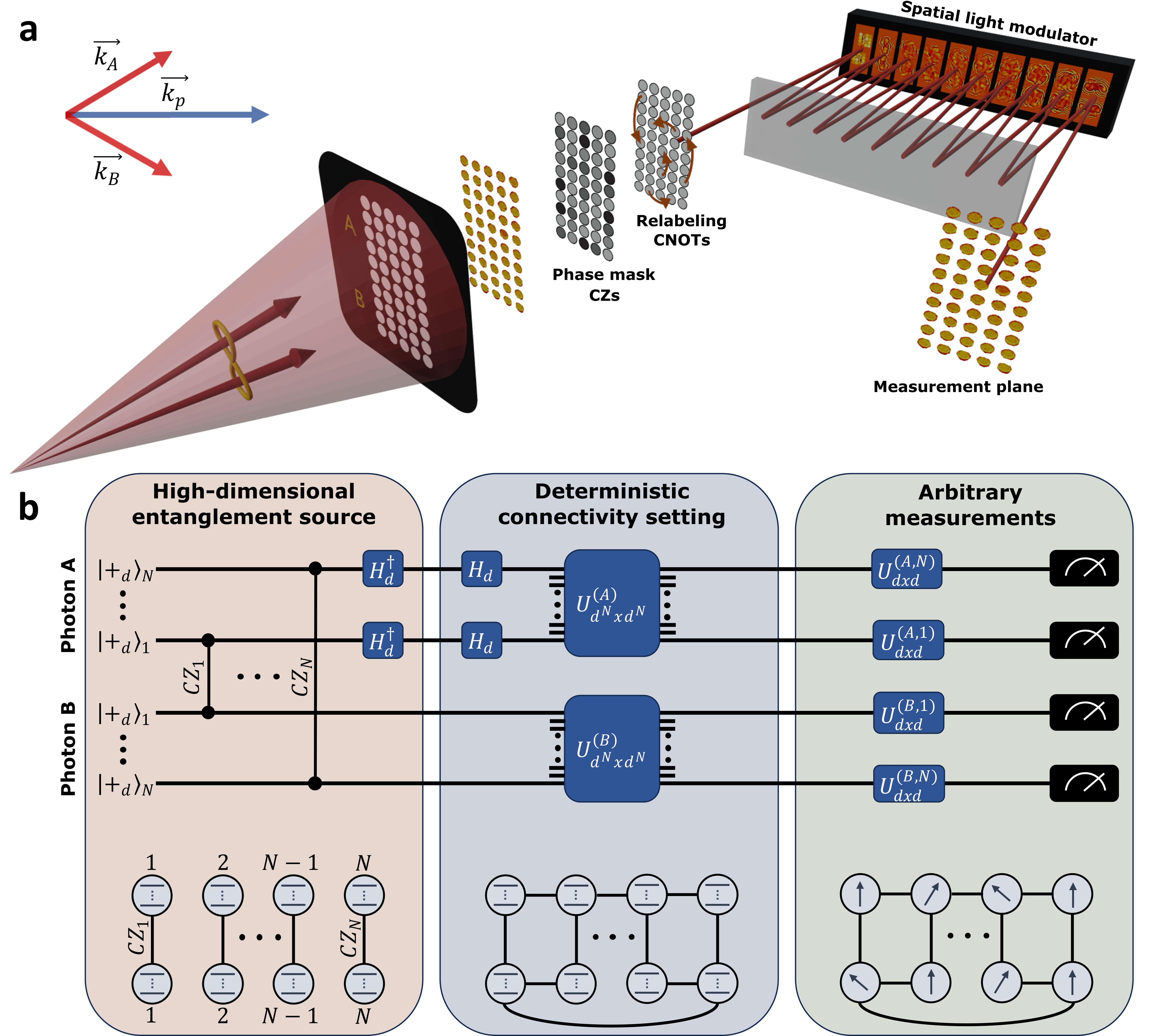}
\caption{\label{fig:1} \textbf{High-dimensional spatially encoded cluster states.} \textbf{a}, Illustration of the experimental setup. Spatially entangled photon pairs are generated via spontaneous parametric down-conversion (SPDC) and are transmitted through a binary mask with a grid of up to 50 apertures. Due to momentum conservation (inset), when photon A passes through one of the top 25 apertures, photon B passes through the anti-symmetric bottom aperture. Mode-dependent phases and mode relabeling are equivalent to entangling CZ and CNOT operations between the qudits encoded in the Hilbert space of each photon, respectively. A multi-plane light converter consisting of a spatial light modulator and a mirror in a multi-pass configuration implements programmable high-dimensional unitary transformations, enabling coincidence measurements in arbitrary bases. \textbf{b}, Quantum circuit with $N$ qudits per photon (top) and qudit graph representations (bottom) of the above experimental setup. The SPDC source generates a state of high-dimensional entangled photons, that is equivalent, up to local Hadamard gates ($H_d$), to a multi-qudit state initialized in $|+_d\rangle^{\otimes2N}=[1/\sqrt{d}(|1\rangle+\cdots+|d\rangle)]^{\otimes2N}$, followed by CZ gates creating pair-wise entanglement between the $2N$ qudits encoded by the two photons (left). Unitary transformations on each photon are equivalent to entangling operations between its encoded qudits (center), allowing for deterministic connectivity settings within the state. Measurement-based quantum computing can then be achieved by measuring the different qudits at arbitrary bases (right).}
\end{figure}

Our scheme for generating high-dimensional cluster states for MBQC is based on the ability to first entangle different photons in high dimensions, then establish entanglement between the multiple qudits encoded within each photon, and finally measure the generated state in different bases. Figure \ref{fig:1} illustrates our implementation of these three tasks on the level of the experimental realization (fig.\ref{fig:1}a), equivalent quantum circuit (fig.\ref{fig:1}b, top), and resulting photonic cluster state (fig.\ref{fig:1}b, bottom). While this scheme is compatible with high-dimensionally and hyper-entangled states encoded in any degree of freedom, our experimental demonstration is the first to use high-dimensionally entangled free-space spatial modes, where tools to precisely control the quantum state, such as Mega-pixel spatial light modulators, are readily available.

First, photon pairs are generated via spontaneous parametric down-conversion (SPDC)\cite{burnham1970observation}, where high-dimensional spatial entanglement in multiple transverse modes is achieved through careful adjustment of the momentum conservation conditions in the nonlinear process\cite{rubin1996transverse,walborn2010spatial} (fig.\ref{fig:1}a, and Methods). A binary amplitude mask defines $M$ spatial modes per photon, which we can use for encoding $N=\log_d(M)$ d-dimensional qudits. The SPDC process is equivalent to the quantum circuit presented on the left panel of fig.\ref{fig:1}b, which generates pair-wise entanglement between qudits encoded in the two photons (fig.\ref{fig:1}b, bottom left). While entangling different photons often relies on probabilistic schemes\cite{calsamiglia2002generalized,paesani2021scheme} and therefore difficult to scale, qudits encoded within the same photons can be arbitrarily entangled through deterministic unitary transformations acting on the high-dimensional Hilbert space of each photon\cite{vallone2007realization,chen2007experimental,reimer2019high,vigliar2021error} (fig.\ref{fig:1}b, center). For example, generalized d-dimensional controlled-Z (CZ) and controlled-NOT (CNOT) gates are implemented in the experiment via mode-dependent phase shifts and mode-relabeling, respectively (fig.\ref{fig:1}a, and supplementary information for additional details). Therefore, we can take advantage of the entanglement generated by the SPDC process (fig.\ref{fig:1}b, left), and deterministically connect qudits encoded on the same photon to create different cluster states using only unitary operations on the modes of each photon (fig.\ref{fig:1}b, center). We discuss the class of cluster states that can be generated in this way in the supplementary information.

After generating the desired cluster state, performing measurements in different bases is essential for both MBQC\cite{raussendorf2003measurement} and entanglement certification\cite{toth2005detecting}. However, while high-dimensional encoding simplifies the generation of large cluster states, it complicates their measurement in different bases. This is due to the increased size of the required unitary transformations that act in the high-dimensional Hilbert space of each photon\cite{wang2018multidimensional}. 

To overcome this challenge, we use multi-plane light conversion (MPLC) - a technique for performing arbitrary transformations on photonic spatial modes by propagating them through a series of optimized phase masks\cite{morizur2010programmable}. While initially developed for transforming spatial modes of classical light\cite{labroille2014efficient,lin2018all,fontaine2019laguerre,kupianskyi2023alloptically}, recent small-scale demonstrations have highlighted the potential of this approach for controlling single and entangled photons\cite{brandt2020high,hiekkamaki2021high,lib2022processing}. We build upon these works to realize the largest programmable multi-plane light converter reported to date consisting of ten programmable phase masks in a multi-pass configuration (fig.\ref{fig:1}a), allowing us to measure all qudits at the required bases (fig.\ref{fig:1}b, right). Importantly, while the number of phase masks required for realizing arbitrary $M$x$M$ unitary transformations scales linearly with $M$, our change-of-basis transformations acting on $N$ qudits consist of a tensor product of $N$, $d$x$d$, unitary matrices acting on each qudit. Therefore, the number of required phase masks scales linearly with $N=\log_d(M)$ and only logarithmically with the number of modes $M$, making this approach more scalable (see supplementary information).

\begin{figure}[H]
\centering
\includegraphics[width=0.66\textwidth]{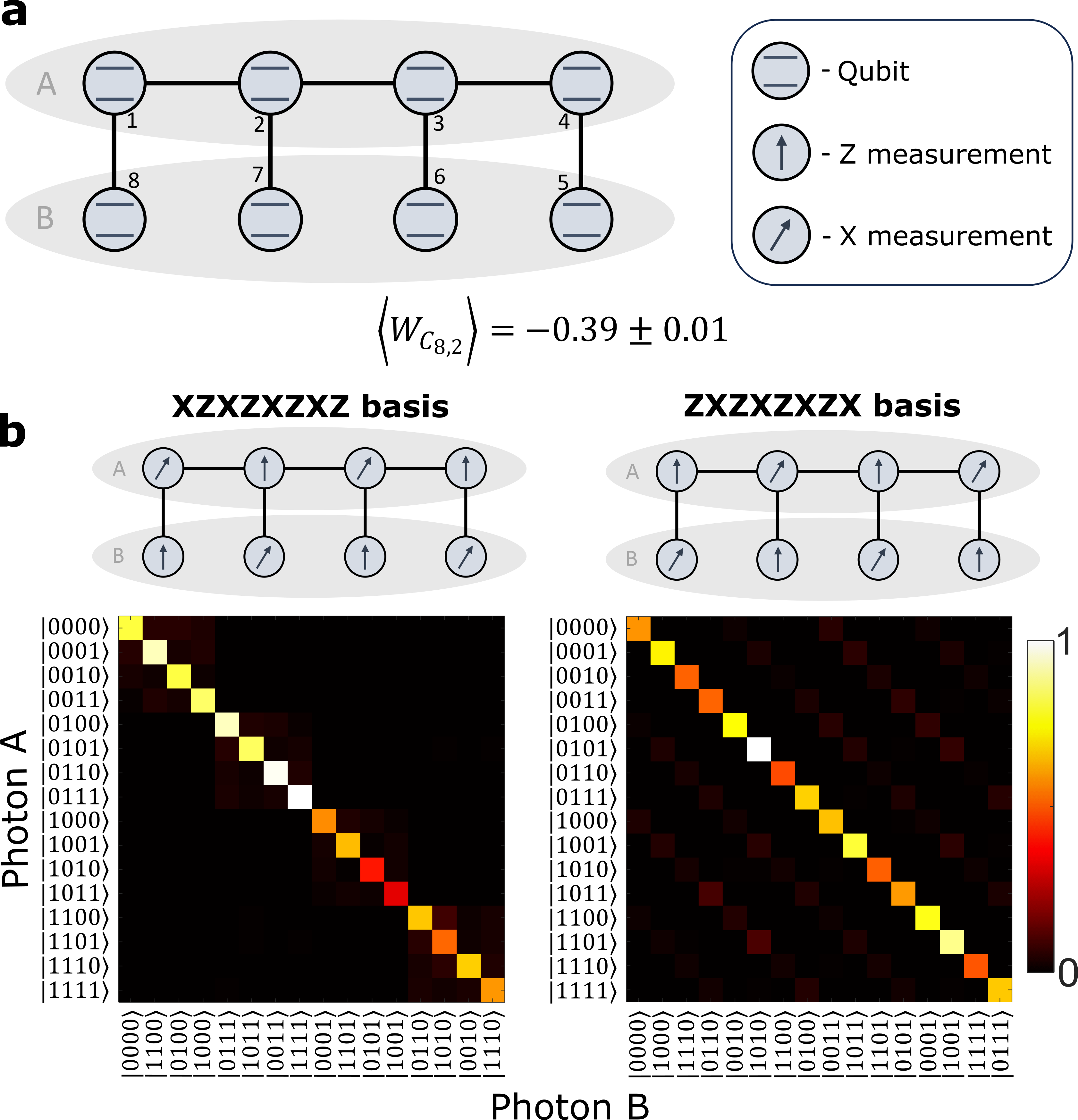}
\caption{\label{fig:2} \textbf{Generation and certification of an eight-qubit cluster state}. \textbf{a}, An eight-qubit cluster state $|C_{8,2}\rangle$ is encoded using two photons (A and B) entangled in $M=16$ spatial modes per photon. Vertices and edges represent qubits and CZ operations, respectively. \textbf{b}, To certify the generation of the desired cluster state, we measure correlations in two mutually unbiased bases consisting of different combinations of Pauli X and Z operators. Based on these measurements, we obtain a negative entanglement witness of $\langle W_{C_{8,2}} \rangle=-0.39\pm0.01$, certifying genuine eight-qubit entanglement.}
\end{figure}

As a first demonstration of cluster state generation using high-dimensional spatial entanglement, we create an eight-qubit cluster state encoded using only two photons (fig.\ref{fig:2}a). In this case, each photon encodes information in $M=16$ spatial modes, that are mapped onto $N=4$ qubits (see supplementary information). By performing two sets of measurements in mutually unbiased bases consisting of different combinations of Pauli X and Z operators, we certify genuine eight-qubit entanglement (fig.\ref{fig:2}b). We observe strong correlations on the diagonal with minor contributions from correlations between other modes interfered via MPLC, obtaining a negative entanglement witness of $\langle W_{C_{8,2}} \rangle=-0.39\pm0.01$\cite{guhne2009entanglement} (see supplementary information). We achieve a total detection rate of over 100Hz in both measurement bases, which is well beyond current capabilities in equivalent eight-photon experiments with two-dimensional encoding\cite{yao2012experimental,thomas2022efficient}.

\subsection*{Efficient MBQC via instantaneous intra-feedforward}

After generating a large cluster state, MBQC is realized via single-qubit measurements that can implement a universal set of quantum gates, such as the one consisting of the two-qubit CNOT gate and general single-qubit rotations\cite{raussendorf2001one,prevedel2007high}. To obtain a deterministic output at the end of the computation, the inherent randomness in the results of the individual single-qubit measurements must be addressed. For Clifford gates, such as the Hadamard or CNOT gates, one could simply keep track of the individual measurement results by updating the so-called Pauli frame according to which the final result is interpreted\cite{raussendorf2003measurement,danos2006determinism}. As single-qubit measurements on different qubits commute, this means that quantum circuits of Clifford gates can be implemented in a single time step in MBQC. However, this is not the case when non-Clifford gates are introduced. For non-Clifford gates, such as single-qubit rotations with rotation angles that are not an integer multiple of $\frac{\pi}{2}$, the measurement bases of the measured qubits must be adapted according to the random measurement results of previously measured qubits. This action of classical feedforward thus sets the temporal order of the computation and, ultimately, its duration. 

When each photon in the cluster state encodes multiple qubits, two different types of feedforward steps can be identified: between qubits encoded on different photons (coined inter-feedforward), and between qubits encoded on the same photon (coined intra-feedforward). While inter-feedforward steps can be realized through active classical hardware as in standard MBQC, the implementation of intra-feedforward steps warrants a different approach. This is because the measurement of a single photon now collapses multiple qubits simultaneously, preventing a naive implementation of feedforward between them. Nevertheless, as we will show, a judicious design of linear optic circuits can take advantage of this collapse for performing multiple feedforward steps instantaneously and passively.

Inspired by early ideas for performing a single feedforward step by measuring the polarization of a photon conditioned on its momentum \cite{vallone2008active}, we show that complete measurement-based gates with multiple intra-feedforward steps can be realized via passive linear optic circuits. We replace active feedforward steps with equivalent multi-qubit gates that reflect the feedforward structure, followed by simultaneous measurement of all qubits encoded in the photon. Such instantaneous feedforward is achieved deterministically through tailored high-dimensional linear optic circuits that implement the multi-qubit gates within the Hilbert space of a single photon. This approach substantially reduces the computation time, as measurements originally made at different time steps with slow, active, feedforward between them, can now be performed simultaneously on qubits encoded in the same photon. We further discuss how this approach can be extended to complete MBQC computation in the supplementary information

\begin{figure}[H]
\centering
\includegraphics[width=1\textwidth]{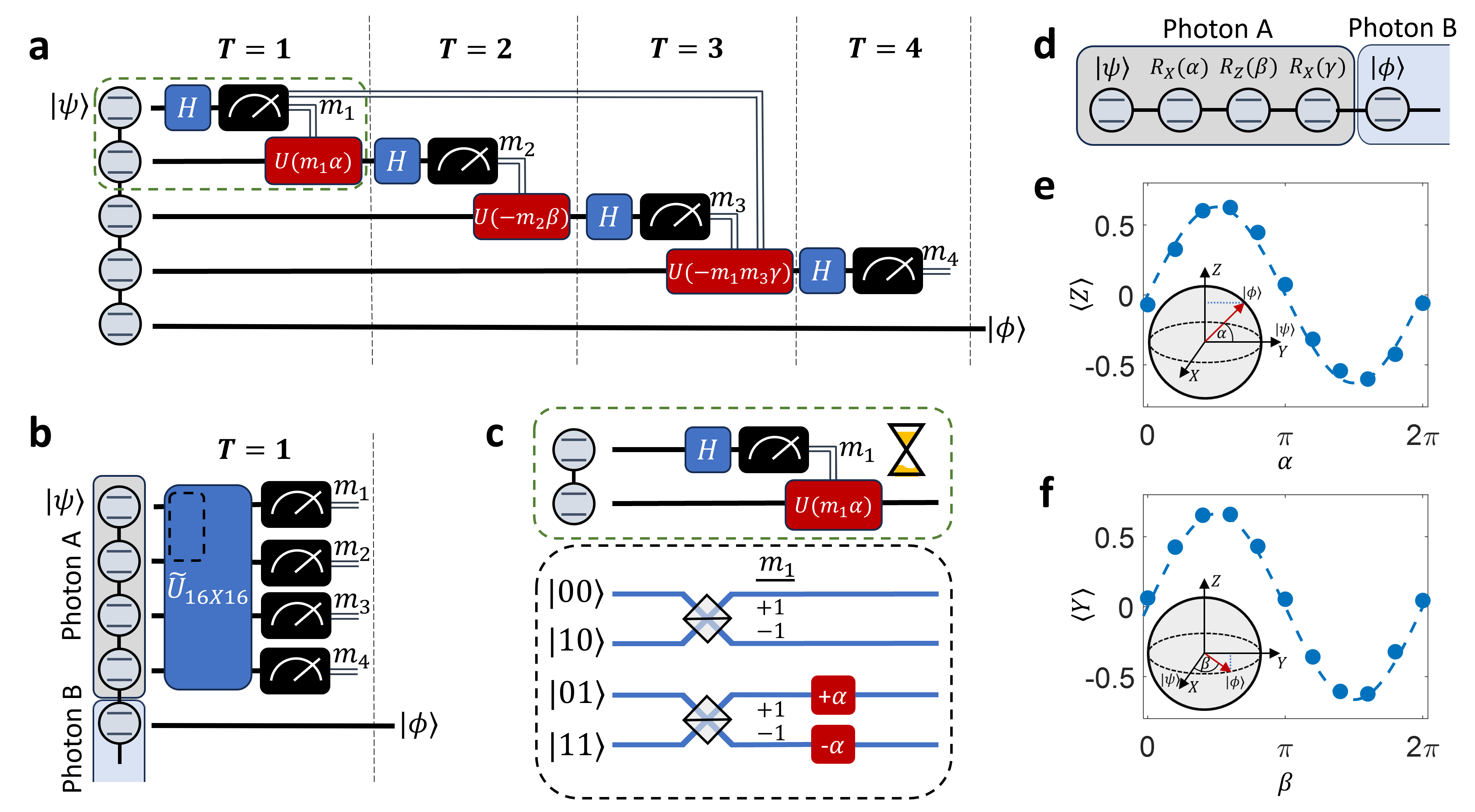}
\caption{\label{fig:3}\textbf{Single-qubit rotation using instantaneous intra-feedforward}. \textbf{a}, In MBQC, arbitrary single-qubit rotations parameterized by rotation angles $(\alpha,\beta, \gamma)$ around the $(X,Z,X)$ axes are performed using a five-qubit chain. Classical feedforward compensates for the inherent randomness in the measurement outcomes $m_i=\pm1$ by changing the measurement basis of the following qubits. This is implemented by a unitary $U(m_i\theta)$ that adds a phase $m_i\theta$ to the $|1\rangle$ state of the qubit. The duration of the total gate is thus dictated by the slow feedforward process that imposes temporal order on the measurements. \textbf{b}, When four qubits are encoded using a single high-dimensional photon, all qubits are measured simultaneously when the photon is detected. Intra-feedforward between them is achieved instantaneously using tailored transformation on the 16-dimensional photon Hilbert space, which can be realized using a 16x16 linear optic circuit. A single intra-feedforward step (top) and its equivalent linear optic circuit (bottom) are shown in panel \textbf{c}. Black and blue lines represent qubits and optical modes, respectively. We experimentally demonstrate single-qubit rotations on a chain of five qubits, four of which are encoded by a single photon (\textbf{d}). The expectation values of the output rotated states $|\phi\rangle$ as a function of angle are presented in panels (\textbf{e}, \textbf{f}). In \textbf{e}, an input state $|\psi\rangle=1/\sqrt{2}(|0\rangle+i|1\rangle)$ is rotated around the X axis by an angle $\alpha$, while in \textbf{f}, an input state $|\psi\rangle=1/\sqrt{2}(|0\rangle+|1\rangle)$ is rotated around the Z axis by an angle $\beta$.}
\end{figure}

We experimentally demonstrate the advantages of instantaneous intra-feedforward for MBQC by implementing measurement-based single-qubit rotations on a five-qubit linear cluster state. While typically non-Clifford single-qubit rotations require four clock cycles and feedforward steps\cite{raussendorf2003measurement} (fig.\ref{fig:3}a), we show how they can be realized in a single time step using intra-feedforward between four qubits encoded on a single photon (fig.\ref{fig:3}b). To this end, we map each slow feedforward step in the original scheme\cite{raussendorf2003measurement} (fig.\ref{fig:3}c, top) with an equivalent linear optical circuit (fig.\ref{fig:3}c, bottom). From these 4x4 circuits, we construct a 16x16 optical circuit that performs all intra-feedforward steps simultaneously, which are then implemented by the MPLC (see fig.S4 in the supplementary information). Following the intra-feedforward steps, the final inter-feedforward step is realized by updating the Pauli frame of the last qubit in the data analysis (see supplementary information). To demonstrate the ability to perform single qubit rotations, we focus on two input states that are rotated by angles $\alpha$ and $\beta$ around the X and Z axes, respectively (fig.\ref{fig:3}d-f, and supplementary information). We observe good agreement of the output expectation values as a function of rotation angle, up to lower visibility resulting mainly from imperfect fidelity, confirming the applicability of instantaneous intra-feedforward for MBQC.

\subsection*{Realization of a $d=5$ qudit cluster state}

Until this point, we have focused on the encoding of multiple \emph{qubits} within each photon in the cluster state. However, high-dimensional encoding naturally provides the flexibility to encode multiple \emph{qudits} as well, as was recently shown for d=3 qudit cluster states\cite{reimer2019high}. While MBQC was predominately studied in the context of qubits, a few notable works have recently suggested potential benefits in implementing gates\cite{zhou2003quantum,gokhale2019asymptotic}, algorithms\cite{wang2017qudit,wang2020qudits,karacsony2023efficient}, and error correction\cite{campbell2012magic} using qudits, especially in prime dimensions\cite{campbell2012magic,booth2023outcome}. 

\begin{figure}[H]
\centering
\includegraphics[width=0.66\textwidth]{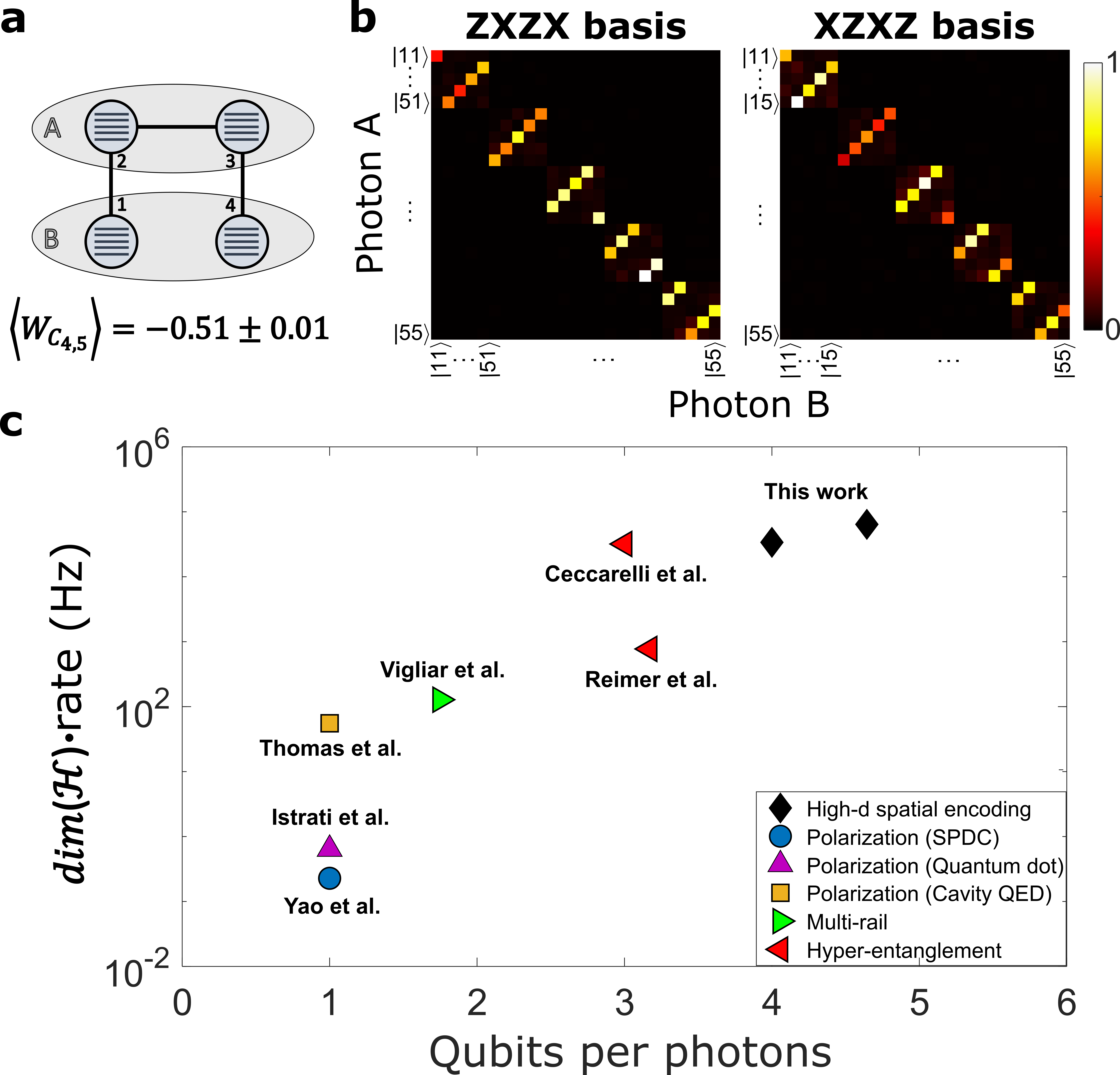}
\caption{\label{fig:4}\textbf{$\mathbf{d=5}$ qudit cluster state}. We certify the generation of a four-partite five-dimensional qudit cluster state (\textbf{a}) via correlation measurements in two bases (\textbf{b}). Here, X and Z are the generalized five-dimensional Pauli operators. We obtain a negative entanglement witness of $\langle W_{C_{4,5}} \rangle=-0.51 \pm 0.01$, certifying genuine four-qudit entanglement. \textbf{c}, The effective quantum resource rate, defined as the product of the state's Hilbert space dimension and detection rate \cite{reimer2019high}, is presented as a function of the number of qubits encoded per photon, for different realizations of photonic cluster states. Our $d=5$ cluster state has a Hilbert space dimension of 625 and a detection rate of approximately 100Hz, yielding a large effective quantum resource rate of dim⁡($\mathcal{H}$)$\cdot$rate $\approx6.25\cdot10^4$Hz. Data for other experiments presented in (\textbf{c}) is taken from fig.\ref{fig:2} (eight qubits at approximately $130$Hz) and the following references \cite{yao2012experimental,ceccarelli2009experimental,reimer2019high,thomas2022efficient,vigliar2021error,istrati2020sequential}.}
\end{figure}

To this end, we have used the same experimental setup to demonstrate a two-photon, $d=5$, four-qudit cluster state encoded using a total of 50 spatial modes (fig.\ref{fig:4}a). To certify genuine entanglement, we use an entanglement witness, adapted for $d=5$ qudit cluster states, that requires measurements in two mutually-unbiased bases\cite{sciara2019universal} (fig.\ref{fig:4}b, see supplementary information). We obtain a negative entanglement witness of $\langle W_{C_{4,5}} \rangle=-0.51 \pm 0.01$, clearly certifying genuine four-qudit entanglement. 

To put our result in perspective, we compare our high-dimensional cluster state with previous demonstrations in terms of both size and detection rate. In terms of size, the Hilbert space dimension of our cluster state ($dim(\mathcal{H})=625$) is equivalent to that of $9.28$ qubits, making it the largest cluster state generated via SPDC\cite{yao2012experimental}. This marks the first time high-dimensional encoding outperforms state-of-the-art SPDC experiments with binary encoding. In terms of rate, our total detection rate of 100Hz is over two orders of magnitude higher than any demonstration of cluster states at this scale, including recent demonstrations with deterministic sources\cite{thomas2022efficient}.

We further quantitatively compare recent two- and high-dimensional demonstrations of cluster states\cite{yao2012experimental,ceccarelli2009experimental,reimer2019high,thomas2022efficient,vigliar2021error,istrati2020sequential} by looking at the effective quantum resource rate as a figure of merit\cite{reimer2019high}. The effective quantum resource rate takes into account both cluster size and detection rate and is defined as the multiplication of the rate by the dimension of the state's Hilbert space. We plot the effective quantum resource rate of different experiments as a function of the number of qubits encoded per photon in figure \ref{fig:4}c, showing that beyond the advantage of instantaneous intra-feedforward, high dimensional encoding is beneficial for high-rate generation of large cluster states. By encoding over four qubits per photon, we are able to achieve a record-high effective quantum resource rate with our $d=5$ cluster state.

\section*{Discussion}
We demonstrate the generation of the largest cluster states ever produced via SPDC, with up to $9.28$ qubits, at high rates. This achievement is made possible by encoding cluster states using high-dimensional spatial encoding, reducing the number of required photons and thus circumventing the size-rate trade-off. We control the large number of photonic modes by implementing a large-scale programmable multi-plane light converter, which allows us to switch between the measurements required for performing measurement-based gates and certifying genuine entanglement. Additionally, we have shown that high-dimensional encoding enables instantaneous feedforward between qubits encoded within the same photon, which substantially boosts the computation speed in MBQC.

This work highlights an important question: What is the optimal encoding dimension for scalable, fault-tolerant, MBQC? Answering this question amounts to finding a sweet spot between the benefits of high-dimensional encoding, such as increased information capacity per photon, and drawbacks, such as complex state generation and manipulation. In our experiment, at least up to 25 modes per photon, the benefit of reducing the number of photons outweighs the added complexity of controlling multiple spatial modes, resulting in substantially higher detection rates compared with two-dimensional encoding.

Our results also emphasize the importance of further research into high-dimensional encoding in deterministic platforms\cite{browne2005resource,raussendorf2007topological,bartolucci2021creation,chen2023heralded,thomas2022efficient,cogan2023deterministic,cao2023photonic}, as an alternative for our SPDC source. Specifically, on the theoretical side, answering open questions on the optimal success probability of heralded generation and fusion of high-dimensional entangled states\cite{calsamiglia2002generalized,joo2007one,luo2019quantum,hu2020experimental,paesani2021scheme} would complement our work and enable a more complete discussion on high-dimensional architectures for fault-tolerant MBQC. On the experimental side, scaling up our demonstration to even larger cluster states would further test the limits of high-dimensional encoding for MBQC. This could be achieved in the near future by increasing the dimension of encoding or the number of high-dimensionally entangled photons used in the experiment.

To this end, the dimension of encoding could be increased through the combination of high-dimensionally entangled photons generated in thin films\cite{okoth2019microscale} and resonant metasurfaces\cite{santiago2022resonant} together with large-scale MPLC\cite{labroille2014efficient,fontaine2019laguerre} or integrated photonic circuits\cite{carolan2015universal,tanomura2022scalable,bao2023very}. Additionally, incorporating state-of-the-art superconducting nanowires or single-photon avalanche diode (SPAD) arrays\cite{morimoto2020megapixel,oripov2023superconducting} with better detection efficiencies will further increase the high detection rate achieved in our experiment. To scale up even further, the number of entangled photons could be increased beyond two using either post-selected or heralded\cite{chen2023heralded,cao2024photonic} schemes, which would increase the size of the cluster state by multiple qudits with the addition of each photon.

In a broader context, we believe that our first demonstration of a high-dimensional cluster state generation scheme that outperforms equivalent qubit experiments will pave the way to new applications in quantum computation and communications in different degrees of freedom of light\cite{fickler2012quantum,kues2017chip,martin2017quantifying,valencia2020high}. In addition to exploring higher rates and larger cluster sizes in MBQC, our scheme holds promise for enhancing long-range quantum communication with cluster-state-based all-photonic quantum repeaters\cite{azuma2015all,buterakos2017deterministic}. After distributing different photons onto different nodes of the quantum network, the number of qubits per node could be increased by encoding multiple qubits per photon. To this end, the cluster states generated in our experiment can be made compatible with long-range free-space or fiber links by leveraging the capability of MPLC to convert hundreds of spatial modes into orbital-angular-momentum or fiber modes\cite{fontaine2019laguerre,lib2022processing}. We thus envision that high-dimensional quantum computing and communications could be combined with recent demonstrations of high-dimensional quantum memories\cite{dong2023highly} to demonstrate important building blocks for a future quantum internet based on high-dimensional encoding\cite{wehner2018quantum,network2023science}.

\bibliography{sn-bibliography}

\section*{Methods}
\subsection*{Experimental setup}
Spatially entangled photons are generated via type-I SPDC by pumping a Barium Borate (BBO) crystal with a $30mW$, $405nm$ continuous-wave laser beam (see figure S1a for a sketch of the experimental setup). The waist of the pump beam at the crystal plane determines the transverse momentum conservation conditions and is inversely proportional to the spatial correlation width between the entangled photons at the far field. The crystal thickness, on the other hand, determines the total width of the far-field intensity distribution. Choosing a pump waist of $w\approx600\mu m$ and crystal thickness of $8mm$, we obtain a correlation width of $\approx 100\mu m$ in a $\approx 6mm$ wide cone at the Fourier plane of a $f=150mm$ lens. We place at the Fourier plane a binary amplitude mask, made of a Chromium-covered glass slide carving 50 circular spatial modes with a radius of $100\mu m$ and a spacing of $300 \mu m$ on a $5X10$ etched grid. The entangled photons are then imaged onto the first plane of the multi-plane light converter (MPLC) using four lenses, having focal lengths $f=50,50,150,150 mm$ respectively. Two mirrors, $M1$, and $M2$, are placed in the image and Fourier planes of the first plane of the MPLC to allow for precise and independent control over the entrance position and angle of the light into the MPLC, respectively. 

The MPLC consists of a spatial light modulator (SLM, Hamamatsu X13138-02), a parallel dielectric mirror placed $43.5 mm$ from the SLM, and a right-angle prism placed $69 mm$ from the SLM (see figure S1b). In total, the entangled photons impinge onto the SLM ten times at different locations before propagating another $43.5 mm$ to two transversely motorized $100 \mu m$ fibers coupled to avalanche photo-diode single-photon detectors. A polarizer and two $810 \pm 10 nm$ filters ensure the detection of photons at the desired polarization and wavelength. Coincidence counts are recorded using a time tagger (Swabian instruments, Time Tagger 20) with a $400 ps$ coincidence window. For the alignment of the MPLC, a flip mirror redirects the light at the output to a camera on a moving rail using an imaging system. The alignment procedure is described in greater detail in the supplementary information.

\subsection*{Calculating phase masks}
For any given transformation, the required phase masks are calculated using the wavefront matching algorithm with 30 iterations, based on the code described in ref. \cite{fontaine2019laguerre}. The size of each phase mask is 140X360 pixels. Crosstalk between pixels and diffraction losses are minimized by restricting the diffraction angles from the phase masks to $15$ percent of the entire angular range set by the $12.5\mu m$ pixels. Once the phase masks are calculated, they are displayed on the SLM without modification. Mode-dependent phases are then added on the first plane of the MPLC to encode the desired cluster state and compensate for small optical aberrations in the preceding optical setup.

\section*{Data availability}
All data is available in the main text and supplementary information or at https://doi.org/10.5281/zenodo.8358012.

\section*{Acknowledgments}
We thank O. Katz, B. Dayan, and Z. Aqua for helpful discussions. This project was supported by the Zuckerman STEM Leadership Program. O.L. acknowledges the support of the Clore Scholars Programme of the Clore Israel Foundation. This research project was financially supported by the State of Lower Saxony, Hannover, Germany. 

\section*{Author contributions}
O.L. and Y.B. conceived and conceptualized the project. O.L. designed and built the experimental setup, carried out the experiment, and performed the data analysis under the supervision of Y.B. All authors contributed to the writing of the manuscript.
 
 \section*{Competing interests}
 The authors declare no competing interests.

\section*{Supplementary materials}
Supplementary Text\\
Figs. S1 to S9\\

\newpage

\def\thefigure{S\arabic{figure}}
\setcounter{figure}{0}

\section{Multi-plane light conversion}
\subsection*{Experimental setup}

\begin{figure}[H]
\centering
\includegraphics[width=0.9\textwidth]{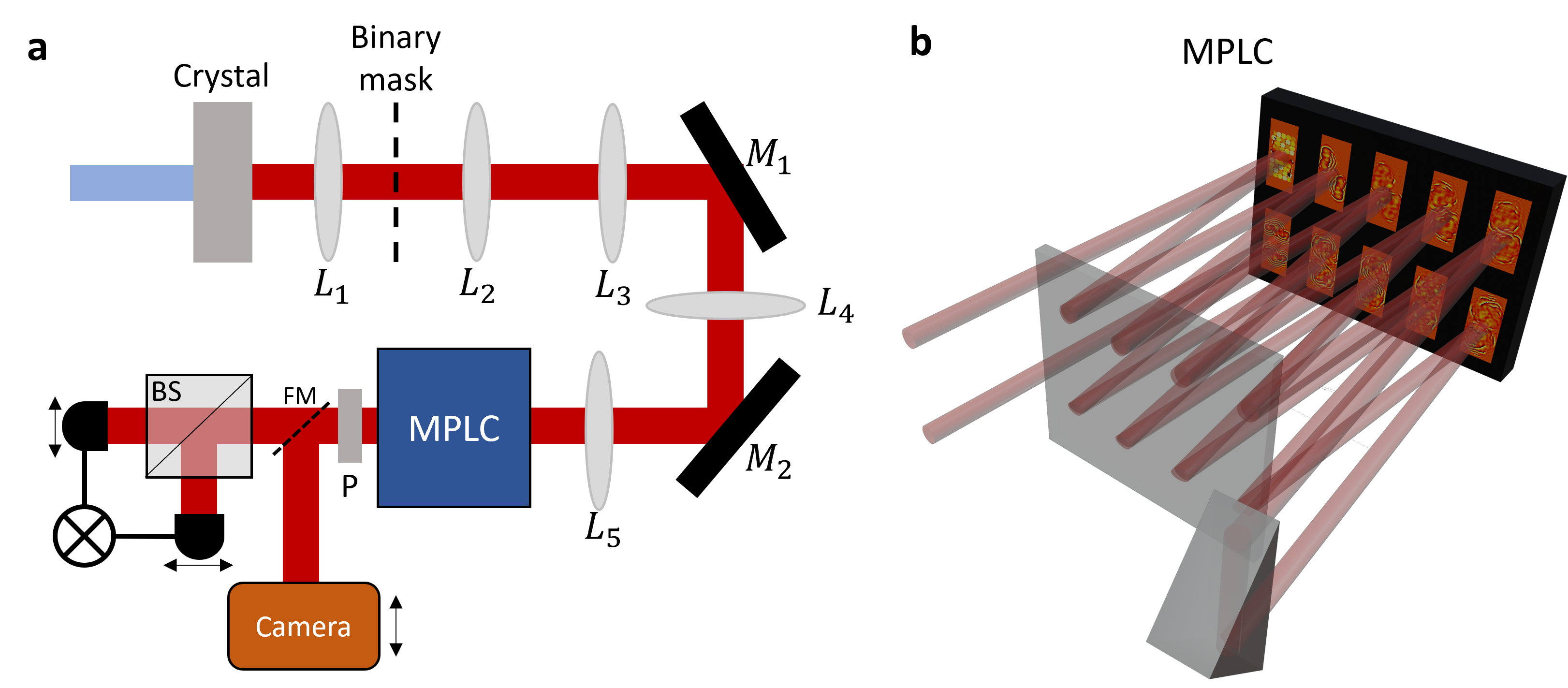}
\caption{\label{fig:S1} \textbf{Experimental setup}. \textbf{a}, A sketch of the experimental setup. $L$- lens, M- mirror, P- polarizer, MPLC- multi-plane light converter, P- polarizer, FM- flip mirror, BS- beam splitter. \textbf{b}, Detailed view of the MPLC. The photons bounce back and forth between a spatial light modulator and a mirror. A prism is used after the fifth plane to redirect the beam onto the lower part of the SLM.}
\end{figure}

The experimental setup is presented in fig.\ref{fig:S1} and is discussed in detail in the Methods section.

\subsection*{Alignment}
Properly aligning MPLCs is known to be experimentally challenging, and is sometimes performed with the aid of optimization algorithms with experimental feedback, even for a relatively small number of planes. In our case, due to the large number of planes and weak SPDC signal, we decided to avoid blind optimization and developed an optical alignment procedure that takes advantage of the large number of planes in our MPLC.

Once the beam is aligned reasonably well and bounces off the SLM ten times, the goal of the alignment procedure is to find the ten $(x_j,y_j), j=1,...,10$ coordinates where the center of the beam hits the SLM. These locations then determine where the phase masks are placed on the SLM. Since an average misalignment of a few pixels is already detrimental to the transformation, we position the phase mask with an accuracy of $\pm1$ pixels. 

Our alignment procedure relies on sequentially imaging the different planes of the SLM onto an sCMOS camera and looking at the intensity distribution of the SPDC light when applying $\pi$ phase steps at different positions on the SLM. The camera is placed on a rail after a set of imaging lenses and can directly image planes nine, ten, and eleven (the effective plane of the detectors) of the MPLC (fig.\ref{fig:S1}a). Planes one to eight are imaged onto the camera as well by adding appropriate quadratic phases to other MPLC planes, which effectively act as programmable imaging lenses. 

The first step of the alignment procedure is to find the position of the beam at the first plane. We image the first plane to the camera and scan the position of a $\pi$ phase step on the SLM. After the imaging system, the $\pi$ phase step is manifested as a dark line on the camera due to high diffraction loss occurring at the phase step. Since each mode at the first plane has a radius of only $100 \mu m$, we were able to find the center of the beam with an accuracy of approximately a single pixel ($12.5 \mu m$). This procedure is repeated for the second plane, albeit with slightly lower accuracy due to the diffraction of the modes.

For planes three to ten, the diffracted SPDC intensity distribution is wide, and it is hard to define the center of the beam. We, therefore, again use other planes of the MPLC as programmable lenses and image the modes from the first plane to the desired plane, and then onto the camera. For example, to find the position of phase mask three, we first image plane one onto plane three by setting a quadratic phase on plane two, and then image plane three onto the camera using a quadratic phase on plane seven. We repeat this process for all planes, including self-consistency checks where the position of certain planes is determined in multiple ways. For example, plane one can be imaged onto plane five using a quadratic phase on either plane three, planes two and four, plane two, or plane four. In addition, planes two and three can be imaged onto plane five as well to check that their measured centers are consistent. The entire alignment procedure takes less than an hour, and the MPLC typically remains aligned and stable for over a month.

\subsection*{Transformation characterization}

To characterize the different transformations implemented via the MPLC, we co-align a classical $808 nm$ laser beam with the SPDC light, through the binary amplitude mask and the MPLC. We reconstruct the matrix $A$ that represents the MPLC transformation using two sets of input modes. First, we inject the laser beam into the MPLC through a single aperture in the binary mask and record the intensities at all output modes. By repeating this measurement for all the apertures, we obtain the absolute values of the matrix elements $|A_{ij}|$, i.e. the square root of the recorded intensity at output mode $i$ for input mode $j$. As, in general, the transformation of the MPLC has finite efficiency (see next section), we do not assume the unitarity of the matrix $A$ at any point. Next, we obtain the relative phases between the matrix elements. We coherently inject light into all the apertures of the binary masks, simultaneously, and look at the output intensities for different relative phases at the input plane (set by the SLM). The relative phases between the matrix elements within each row of the matrix can then be obtained through a simple implementation of the Gerchberg-Saxton phase retrieval algorithm. Obtaining the relative phases within each row is sufficient since the relative phases between rows, or equivalently between output modes, are irrelevant to our intensity and correlation measurements at the output.

We experimentally reconstructed the 18 transformations used for evaluating the entanglement witnesses in figures 2,4. For the eight-qubit cluster states, the desired transformations on each photon are four-dimensional Hadamard operations $U=H_4$ between 8 different sets of four modes, while for the four-qudit cluster state, five-dimensional discrete Fourier transformation $U=DFT_5$ between 10 sets of modes are required. We compare the reconstructed matrices $A$ with the desired ones $U$ to obtain their fidelity according to the Frobenius norm, $F=\frac{|Trace(AU^\dag)|}{\sqrt{Trace(AA^\dag)Trace(UU^\dag)}}$, achieving $F_{H_4}=0.95\pm0.02$ and $F_{DFT_5}=0.93\pm0.01$.

\subsection*{Loss estimation}
To estimate the optical loss in the system, we measured the total coincidence rate before the MPLC and compared it with that from the entanglement certification measurements of the eight-qubit and four-qudit cluster states after the MPLC. The same detectors, fibers, and filters are used for all measurements. As the two-photon coincidence rate drops quadratically with loss, the estimated loss is given by the square root of the ratio between the output and input coincidence rates. The measured average losses for the eight-qubit and four-qudit cluster states are $-9.1 \pm 0.1 dB$ and $-10.4 \pm 0.2 dB$, respectively. The use of high-dimensional encoding enables us to obtain high detection rates of large cluster states in the presence of such loss. The detection rate could be boosted even further by using fixed phase plates instead of an SLM, which can substantially reduce loss.

\section{Encoding cluster states with high-dimensional entanglement}

In this section, we describe the mapping between two-photon high-dimensional entangled states generated via SPDC and multi-qubit/qudit graph states. Due to transverse momentum conservation, for $M$ modes per photon, the output state of the entangled photons is $|\psi\rangle = \frac{1}{\sqrt{M}} \sum_{i=1}^{M} |ii\rangle$, where modes that are anti-symmetric with respect to the center of the beam have the same label $i=1,...,M$. For any given total dimension $M$, the state of each photon can be represented using $N$, $d$-dimensional, qudits with $d^N=M$. The two-photon quantum state can thus be written as a sum over all possible strings $s_i$ of $N$ numbers from 1 to $d$, $|\psi\rangle = \frac{1}{\sqrt{d^N}} \sum_{s_i=\{1,...,d\}^N} |s_is_i\rangle$. Note that all possible strings for each photon are represented in the sum and that the value of qudit $n\in\{1,...,N\}$ in both photons is always identical due to their entanglement. Therefore, the quantum state is equivalent to a tensor product of $N$, $d$-dimensional, entangled states and takes the form $|\psi\rangle=(\frac{1}{\sqrt{d}} \sum_{i=1}^{d} |ii\rangle)^{\otimes N}$. This state is equivalent to the graph state presented on the left side of fig.1b, up to the local operation of the conjugate transpose of the $d$-dimensional Hadamard gate ($H_d^\dag$) on each of the qudits on one of the two photons (figure 1b, left).

\begin{figure}[H]
\centering
\includegraphics[width=0.75\textwidth]{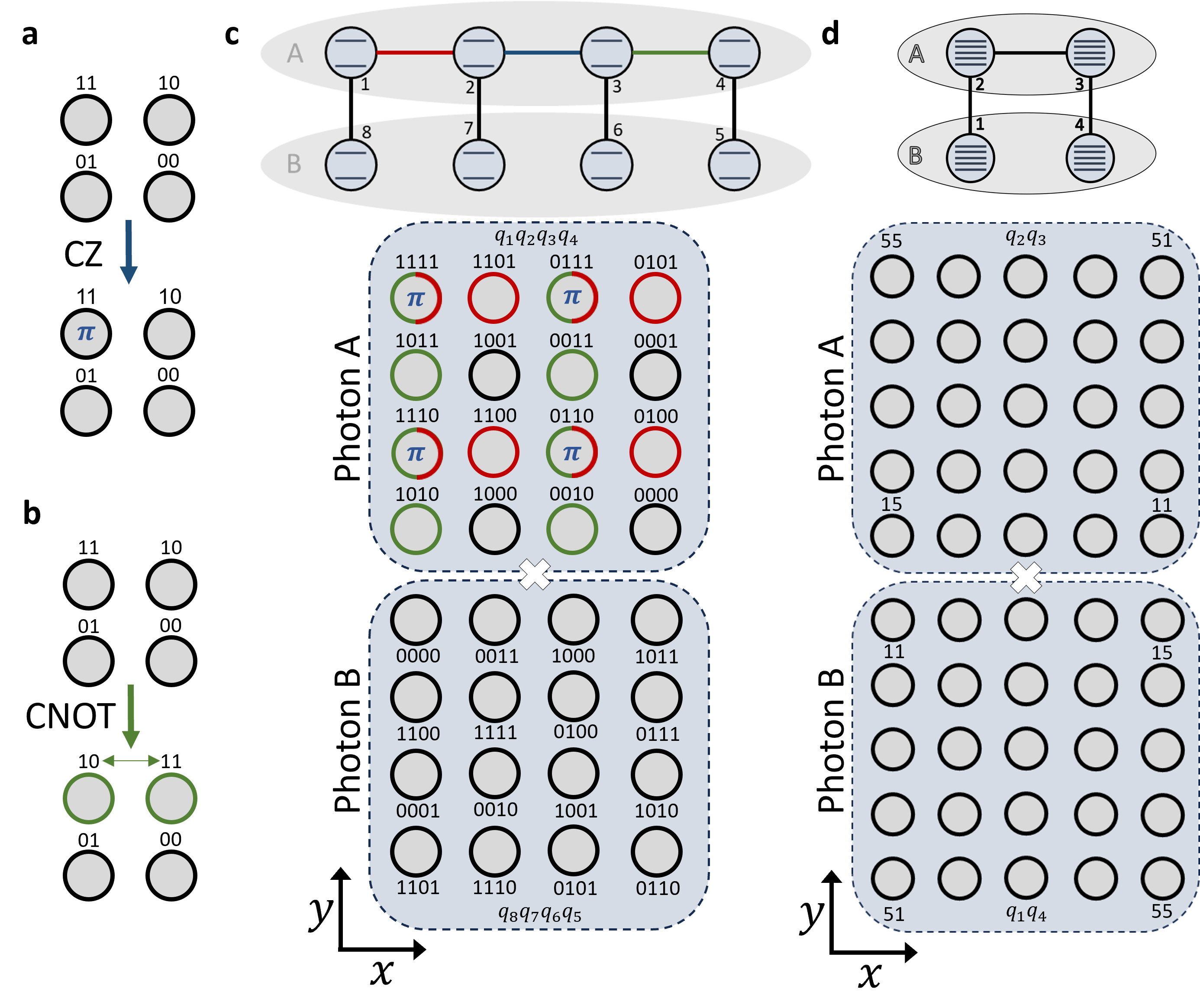}
\caption{\label{fig:S2} \textbf{Cluster state encoding}. \textbf{a}, A CZ gate between two qubits is equivalent to a $\pi$ phase shift on the spatial mode encoding the $|11\rangle$ state. \textbf{b}, Similarly, a CNOT gate between two qubits is equivalent to the label swapping of the modes encoding the $|10\rangle$ and $|11\rangle$ states (outlined in green). \textbf{c}, The eight-qubit cluster state is encoded in 16 spatial modes per photon. The modes of the two photons are positioned in anti-symmetric locations with respect to the center of the 32 modes, marked with a white X sign. Edges in the graph representation of the state are realized via mode-dependent phase shifts (CZ) and mode-relabeling (CNOT). The blue edge is equivalent to a $\pi$ phase shift added to modes where both the second and third qubits have the value $1$. The red and green edges are equivalent to mode relabeling (see text above). Modes that have been relabeled in the addition of the red and green edges are marked with the corresponding color. \textbf{d}, The four-qudit $d=5$ cluster state is encoded using 25 spatial modes per photon. For clarity, only the labels 11,15,51, and 55 are explicitly written in the figure. The edge between qudits two and three is equivalent to the addition of a phase $(i-1)(j-1)2\pi/5$ to each mode $ij$ of photon A.}
\end{figure}

After the SPDC process creates pair-wise entanglement between the qudits of the two photons (fig.1B, left), the connectivity of the cluster state is set by entangling qudits encoded within the same photon using local unitary transformations on the optical modes. CZ gates can be directly applied to qudits that did not experience the extra $H_d^\dag$ gate. The CZ gate between two qudits encoded on the same photon is equivalent to a mode-dependent phase shift of $(i-1)(j-1)2\pi/d$ where $i=1,...,d$ and $j=1,...,d$ are the values of the two qudits at this mode. For example, in the qubit case, this amounts to a $\pi$ phase shift of the $11$ mode (considering standard 0/1 encoding of the qubits), as can be seen in fig.\ref{fig:S2}a. To apply a CZ gate between two qudits where one of them has an additional $H_d^\dag$ gate, the CZ gate is sandwiched between two $H_d$ gates, making it equivalent to a CNOT gate. The first $H_d$ gate compensates for the extra $H_d^\dag$ gate resulting from the SPDC process while the second one is absorbed in the measurement basis. The CNOT gate between two qudits encoded on the same photon is equivalent to relabeling. For example, in the qubit case, a CNOT gate between two qubits is equivalent to the label swapping of the $11$ and $10$ modes (fig.\ref{fig:S2}b).

For the eight-qubit cluster state, each photon encodes $N=4$ qubits ($d=2$) in $M=16$ optical modes. We choose the order of the labels to be $q_1q_2q_3q_4$ for photon A and $q_8q_7q_6q_5$ for photon B (fig.\ref{fig:S2}c). When choosing identical labels for anti-symmetric modes with respect to the center of the 32 apertures, the SPDC process ensures entanglement between the qubit pairs 1-8, 2-7, 3-6, and 4-5. Following the SPDC process, a CZ gate is applied to qubits two and three, and a CNOT gate between qubits one and two, and between qubits three and four (see fig.\ref{fig:S2}c). The CZ gate is equivalent to a $\pi$ phase shift on modes where $q_2q_3=11$, and the CNOT gates are equivalent to mode relabeling, as shown in fig.\ref{fig:S2}b. Specifically, the CNOT gate between qubits one and two (marked with red) is equivalent to the label swapping of modes where $q_2=1$, by changing $q_1$ from 0 to 1 or vice versa. Similarly, the CNOT gate between qubits three and four (marked with green) is implemented by changing the label of $q_4$ whenever $q_3=1$. The effect of the CZ and CNOT gates is illustrated in fig.\ref{fig:S2}c. 

Similarly, for the four-qudit cluster state, each photon encodes $N=2$, $d=5$, qudits in $M=25$ spatial optical modes. A CZ operation between qudits two and three is performed by applying mode-dependent phase shifts to form the four-qudit cluster (up to Hadamard operations on qudits one and four, fig.\ref{fig:S2}d). In this case, a phase shift of $(i-1)(j-1)2\pi/d$ is applied for a mode labeled with $q_2q_3=ij$.

\section{Entanglement certification}

To experimentally certify the entanglement in the generated cluster states, we use entanglement witnesses $W_{2N,d}$, which are evaluated by measuring the correlations between the two photons (each encoding $N$ qudits) in two mutually unbiased bases. Measurements in two bases are sufficient in our case, as the graphs associated with the quantum state are two-colorable\cite{guhne2009entanglement}. For the eight-qubit case, we use a standard entanglement witness of the form $W_{2N,2}=3-2(\prod_{odd\, k}\frac{(S_k+1)}{2}+\prod_{even\, k}\frac{(S_k+1)}{2})$, where $S_k$ is the stabilizer associated with the kth qubit, given by the Pauli $X$ operator acting on the kth qubit and Pauli Z operator acting on its neighbors\cite{guhne2009entanglement}. Similarly, for the four-qudit cluster state, we use a recent generalization of this entanglement witness that takes the form $W_{2N,d}=\frac{d+1}{d-1}-\frac{d}{d-1}(\prod_{odd\, k}(\frac{1}{d}\sum_{p=1}^{d}S_k^p)+\prod_{even\, k}(\frac{1}{d}\sum_{p=1}^{d}S_k^p))$, where $d$ is the dimension of the qudit and $S_k^p$ is the stabilizer $S_k$ in power $p$\cite{sciara2019universal}. 

In the main text, for clarity, the measurement bases are written according to the standard cluster state stabilizers, absorbing into the measurement the additional local Hadamard gates that act on some of the qubits. Here, we write the stabilizers and entanglement witnesses for the quantum state, including the local Hadamard gates, to make the actual measurement bases realized via the MPLC in the experiment more transparent. The entanglement witnesses for the eight-qubit and four-qudit cluster states are given by

\begin{equation}
   \begin{array}{l} W_{C_{8,2}}=3-\frac{1}{8}( X_2 Z_3 Z_4 Z_5 Z_6 X_7 X_8 +X_2 Z_6 X_7 X_8 + X_2 Z_4 Z_5 X_7 X_8+ X_2 Z_3 X_7 X_8+X_1 Z_4 Z_5 Z_6 X_8\\ 
+ X_1 Z_3 Z_6 X_8 +X_1 Z_3 Z_4 Z_5 X_8+ X_1 X_8+X_1 X_2 Z_3 Z_4 Z_5 Z_6 X_7+X_1 X_2 Z_6 X_7 +X_1 X_2 Z_4 Z_5 X_7\\
+X_1 X_2 Z_3 X_7+Z_4 Z_5 Z_6+Z_3 Z_6+Z_3 Z_4 Z_5+Z_1 Z_2 X_3 X_5 X_6 Z_7 Z_8+ Z_1 Z_2 X_3 X_4 X_6 Z_7 Z_8+ Z_1 X_4 X_5 Z_7 Z_8 \\
+Z_1 Z_7 Z_8+Z_1 X_3 X_5 X_6 Z_8+Z_1 X_3 X_4 X_6 Z_8+Z_1 Z_2 X_4 X_5 Z_8+Z_1 Z_2 Z_8+X_3 X_5 X_6 Z_7+X_3 X_4 X_6 Z_7 \\
+Z_2 X_4 X_5 Z_7+Z_2 Z_7\left.+Z_2 X_3 X_5 X_6+Z_2 X_3 X_4 X_6+X_4 X_5+2\right)

\end{array}
\end{equation}

\begin{equation}
   \begin{array}{l} W_{C_{4,5}}=\frac{14}{10}-\frac{1}{10} Re(Z_1^4 Z_2^2 X_3 X_4+Z_1^4 Z_2^3 X_3^2 X_4^2+Z_1^4 Z_2^4 X_3^3 X_4^3+Z_1^4 I_2 X_3^4 X_4^4 \\
   +Z_1^3 Z_2^3 X_3 X_4+Z_1^3 Z_2^4 X_3^2 X_4^2+Z_1^3 I_2 X_3^3 X_4^3 +Z_1^3 Z_2 X_3^4 X_4^4+I_1 Z_2 X_3 X_4\\
 +I_1 Z_2^2 X_3^2 X_4^2+Z_1^4 Z_2 I_3 I_4+ Z_1^3 Z_2^2 I_3 I_4+I_1 I_2 Z_3 Z_4^4+I_1 I_2 Z_3^2 Z_4^3+X_1 X_2 Z_3^2 Z_4^4 \\
+X_1 X_2 Z_3^3 Z_4^3+X_1 X_2 Z_3^4 Z_4^2+X_1 X_2 I_3 Z_4+X_1 X_2 Z_3 I_4+X_1^2 X_2^2 Z_3^3 Z_4^4+X_1^2 X_2^2 Z_3^4 Z_4^3 \\
\left.+X_1^2 X_2^2 I_3 Z_4^2+X_1^2 X_2^2 Z_3 Z_4+X_1^2 X_2^2 Z_3^2 I_4\right).

\end{array}
\end{equation}

\begin{figure}[H]
\centering
\includegraphics[width=0.9\textwidth]{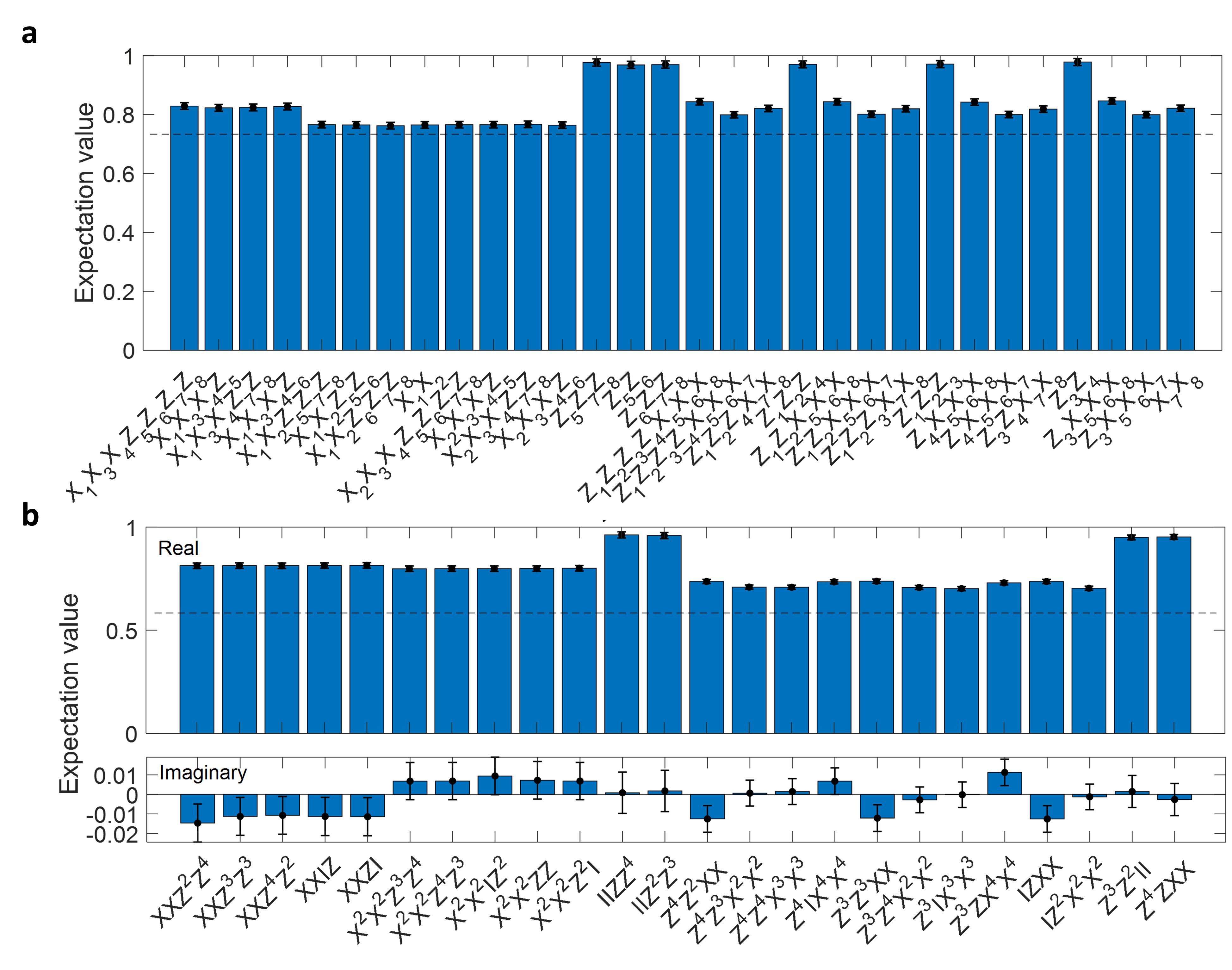}
\caption{\label{fig:S3} \textbf{Entanglement certification}. Expectation values for all terms in the entanglement witness of the eight-qubit (\textbf{a}) and four-qudit (\textbf{b}) cluster states. All values are above the required average to obtain a negative entanglement witness (black dashed line). In the qudit case, the individual expectation values can have imaginary components, which are expected to be zero and are shown to indeed be negligible. The black error bars correspond to the standard deviation calculated from the Poissonian noise in the coincidence counts.}
\end{figure}

For the eight-qubit cluster state, each measurement basis includes two Pauli $X$ operators for each photon. This is because there are at most two Pauli $X$ operators per photon in each term of the entanglement witness, and when there are fewer, they are replaced with Identity operators, which can be evaluated in the $X$ basis as well. The MPLC is thus used to perform the tensor product of two Hadamard operations on the modes of each photon, interfering different sets of four spatial modes. Similarly, for the four-qudit cluster state, each measurement basis consists of at most one single five-dimensional Pauli $X$ operator per photon, and the MPLC performs the five-dimensional discrete Fourier transform, interfering different sets of five spatial modes. In this case, one measurement basis amounts to interfering the modes over the rows of the 5x5 mode grid and the other over the columns. The expectation values of all terms in the expressions for the entanglement witnesses are presented in fig.\ref{fig:S3}, clearly showing genuine entanglement in the generated states.

\section{Implementation of instantaneous feedforward}

\subsection*{Obtaining the five-qubit linear cluster state}

For the demonstration of single-qubit rotation, we use a linear cluster state with four qubits encoded on photon A and a fifth qubit encoded on photon B (fig.3d). This state is obtained from the eight-qubit cluster state by appropriately measuring qubits six to eight. Qubits six and seven are removed from the cluster by projecting them onto the $+1$ eigenstate of the $Z$ basis, which is equivalent to the $X$ basis in the experiment due to the additional Hadamard gates acting on these qubits. Qubit eight is measured in either the $Z$ or $Y$ basis, to both remove it from the cluster and to change the input state $|\psi\rangle$ on qubit $1$.

\subsection*{Instantaneous feedforward with linear optics}
\begin{figure}[H]
\centering
\includegraphics[width=0.9\textwidth]{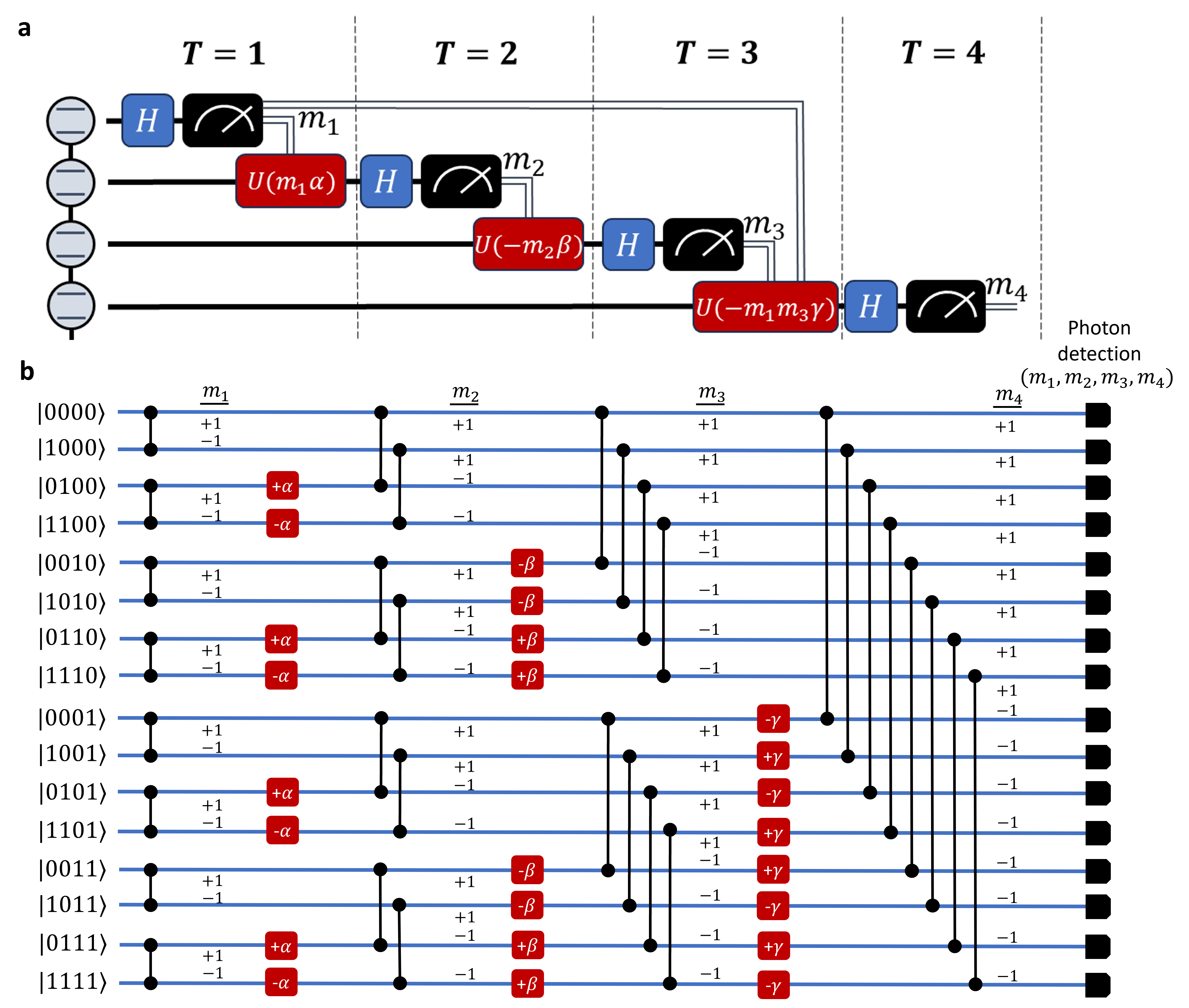}
\caption{\label{fig:S4} \textbf{Instantaneous feedforward with linear optics}. In MBQC, a general single-qubit rotation is performed by measuring four qubits out of a five-qubit chain (\textbf{a}). Feedforward between measurements is used for counteracting the inherent randomness of the measurement results. When the four qubits are encoded within the same photon, as is the case in our experiment, an equivalent linear optic circuit acting on sixteen optical modes can be used instead (\textbf{b}). In \textbf{b}, red rectangles represent phase shifters and black vertical lines with end dots represent beam splitters.}
\end{figure}

As described in the main text, single qubit rotations in MBQC are implemented using a chain of five qubits. In standard binary encoded MBQC, the input state on the first qubit is rotated and teleported to the fifth qubit by measuring qubits one to four. The rotation is parameterized by three rotation angles, $(\alpha,\beta,\gamma)$, around the $(X,Z,X)$ axes. To counteract the randomness in the measurement results of the four qubits, their measurement is performed sequentially in bases determined by previous measurement results (fig.\ref{fig:S4}a). Using this measurement structure with classical feedforward, the output state equals the desired state up to known Pauli correction dependent on the results of the previous measurements.

In our experiment, the first four qubits are encoded using sixteen spatial modes of a single photon. Since the qubits are encoded using a single photon, multi-qubit gates that mimic the structure of the standard quantum circuit with classical feedforward can be implemented using a linear optics circuit encoding the desired unitary transformation on the optical modes. To this end, we designed a linear optic circuit that performs intra-feedforward between four qubits encoded on a single photon for measurement-based single-qubit rotations, as described in fig.\ref{fig:S4}b. For single qubit rotation, every qubit is measured on the equatorial plane of the Bloch sphere, at some angle $\theta$ determined by the rotation angles and the measurement results of previous qubits. We directly implement this in linear optics using phase shifters with an angle $\pm\theta$ added to the relevant modes, followed by a set of beam-splitters. As the structure of the linear optic circuit is independent of the measurement results and the specific angles of the single-qubit rotation gate, it can efficiently be designed in advance before running the computation. This approach circumvents the need for the slow feedforward steps between the measurement of different qubits and enables fast single-qubit rotations without the use of active feedforward between qubits encoded on a single photon. The known Pauli corrections on the last qubit are taken into account in the data analysis. As we discuss in the section ‘Scaling instantaneous intra-feedforward to complete quantum circuits’ of the supplementary information, this approach can be extended to complete quantum circuits in a similar manner.

Specifically, in our implementation, we set $\gamma=0$ and start from an eight-qubit cluster state, as described in the previous section. The first and last layers of eight beam splitters in fig.\ref{fig:S4}b are automatically realized by the extra Hadamard gates on qubits one and four in the generation of the state. The rest of the linear optic circuit is realized via the MPLC. We add the mode-dependent phases $\pm\alpha$ on the first MPLC plane, followed by the relevant beam splitter operations simultaneously realized via planes one to six of the MPLC. The phases $\pm\beta$ are then added on plane six, followed again by eight simultaneous beam splitter operations realized via planes six to ten of the MPLC. For the second photon, qubits six and seven are measured in the $X$ basis as described before. Qubits five and eight are measured at different bases depending on the prepared input state and the output measurement basis.

\subsection*{Extended results}

\begin{figure}[H]
\centering
\includegraphics[width=0.9\textwidth]{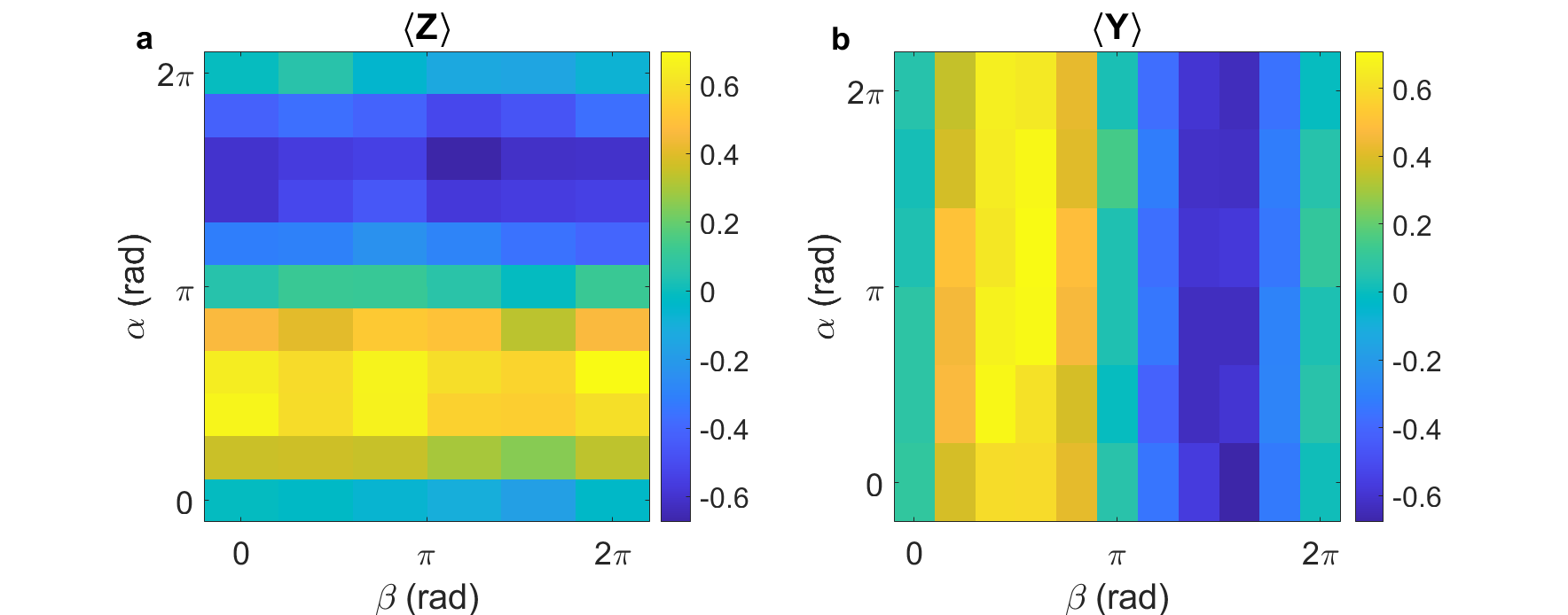}
\caption{\label{fig:S5} \textbf{Single qubit rotation}. Expectation values for the Pauli $Z$ (\textbf{a}) and $Y$ (\textbf{b}) operators as a function of the rotation angles $\alpha$ and $\beta$, for input states $|\psi\rangle=\frac{1}{\sqrt{2}}(|0\rangle+i|1\rangle)$ and $|\psi\rangle=\frac{1}{\sqrt{2}}(|0\rangle+|1\rangle)$, respectively.}
\end{figure}

We demonstrate measurement-based single-qubit rotations by looking at the oscillations of output expectation values of the rotated state as a function of the rotation angles $\alpha$ and $\beta$ around the $X$ and $Z$ axes, respectively. We choose two different sets of an input state and output observable to demonstrate dependence on both $\alpha$ and $\beta$ (fig.\ref{fig:S5}). Despite reduced visibility resulting from noise and imperfect transformations, good agreement with the expected rotations is observed in both cases. The results in fig.3e,f of the main text are one-dimensional averages over the data in fig.\ref{fig:S5}a,b, respectively. To reduce the measurement time in the single-qubit rotation experiment, in this set of measurements, we raised the power of the pump power to 125mW.

\section{Outlook on the realization of complete quantum circuits}

In the experiment, we show that encoding multiple qubits onto each photon may increase the generation rate of cluster states and reduce the feedforward overhead for implementing a non-Clifford gate. In this section, we discuss how the ideas presented in our paper can be generalized in the future to larger cluster states, benefiting implementations of complete MBQC circuits.

\subsection*{Generation of different graph states}

The compatibility of graph and cluster states with different applications highly depends on their structure. For example, it is known that one-dimensional cluster states only allow for arbitrary manipulation of a single qubit, while two- and three-dimensional ones enable universal and fault-tolerant quantum computation, respectively\cite{briegel2009measurement}. Other graph states, such as the ring-shaped graph state or tree cluster states, have found applications in fusion-based quantum computation\cite{bartolucci2023fusion} and quantum repeaters\cite{buterakos2017deterministic}. 

When each photon encodes a single qubit, the structure of the generated cluster state is mainly determined by the physical process that generates the entanglement between photons. This has limited recent demonstrations using cavity-QED and quantum dots to one-dimensional cluster states\cite{istrati2020sequential,cogan2023deterministic,thomas2022efficient}, which require further fusion operations to be extended to more complex topologies\cite{thomas2024fusion}. 

\begin{figure}[H]
\centering
\includegraphics[width=0.9\textwidth]{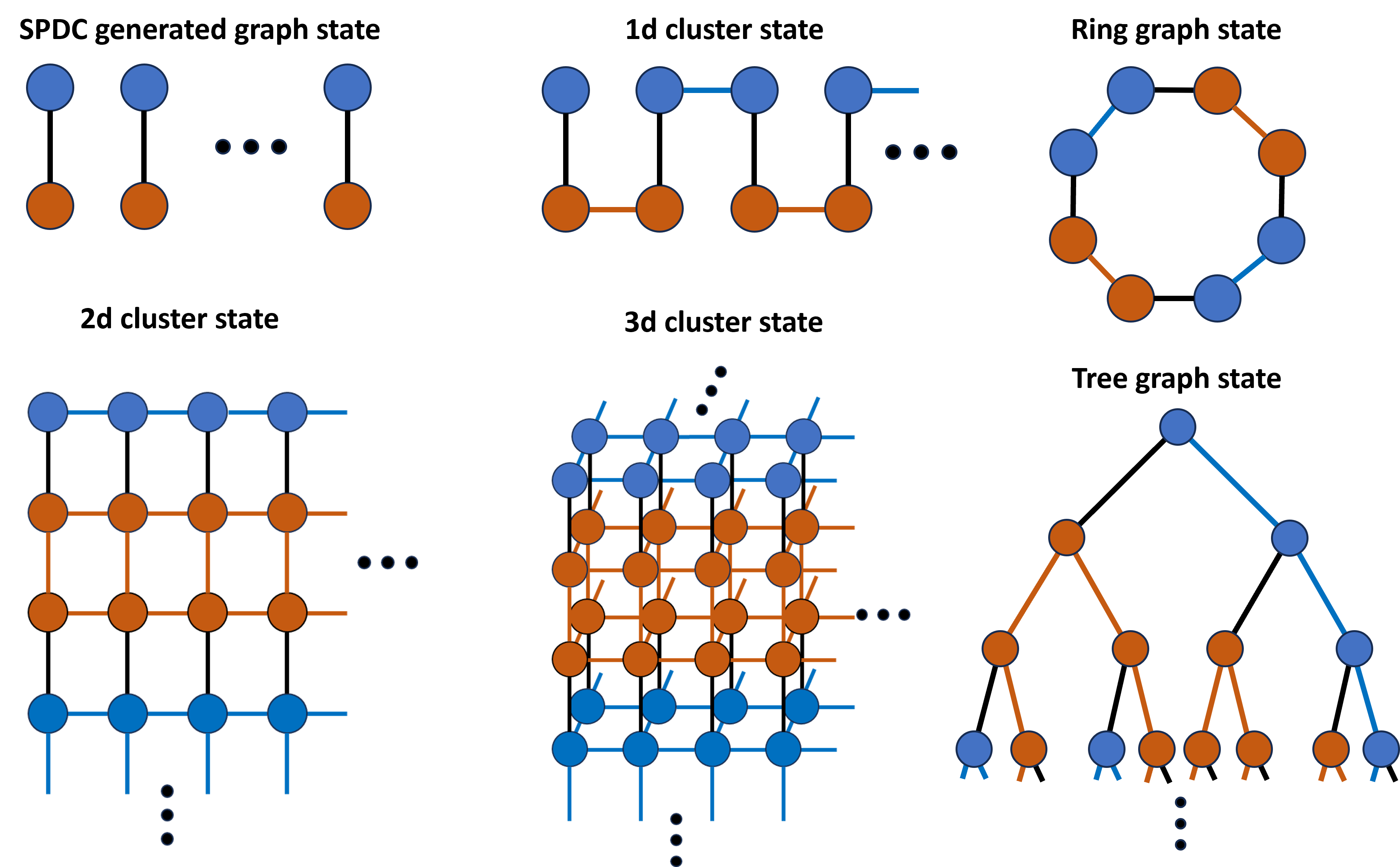}
\caption{\label{fig:S6} \textbf{Generation of different graph states.} Blue and orange vertices represent qubits (or qudits) encoded by the two different photons. The SPDC process creates black edges between pairs of vertices with a different color, while blue and orange edges can arbitrarily connect blue and orange vertices, respectively. Different graph states can be constructed by laying out their desired structure and then starting coloring the vertices and edges one by one according to the following rules: (1) each vertex is connected to exactly one vertex of a different color with a single black edge. (2) vertices of the same color can be arbitrarily connected with an appropriately colored edge.}
\end{figure}

When multiple qubits are encoded onto each photon, the entanglement between photons is again set by the physical entangling process, but the entanglement connectivity between qubits encoded onto the same photon can now be arbitrarily programmed using linear optics\cite{vigliar2021error}. This is the case in our experiment, where the SPDC process generates entanglement between the associated qubits of the two photons (fig.\ref{fig:S6}, top left), and the connectivity of the graph state between qubits encoded on the same photon is easily set, for example, using mode-dependent phase shifts. The class of graph states that can be generated using our experimental approach is thus described by the following adjacency matrix:

\begin{equation}
\begin{pmatrix}
A_{NXN} & \begin{matrix} 1 & 0 & \cdots & 0 & 0 \\ 0 & 1 & 0 & \cdots & 0
\\ 0 & 0 & \ddots & 0 & 0
\\ 0 & \cdots & 0 & 1 & 0
\\ 0 & 0 & \cdots & 0 & 1 \end{matrix} \\
\begin{matrix} 1 & 0 & \cdots & 0 & 0 \\ 0 & 1 & 0 & \cdots & 0
\\ 0 & 0 & \ddots & 0 & 0
\\ 0 & \cdots & 0 & 1 & 0
\\ 0 & 0 & \cdots & 0 & 1 \end{matrix} & B_{NXN}
\end{pmatrix}
\end{equation}

where the SPDC process determines the off-diagonal blocks, and $A_{NXN}, B_{NXN}$ are arbitrary adjacency matrices set by the linear optic circuit applied to the first and second photons, respectively.

This set of graph states can also be understood pictorially by coloring qubits encoded onto different photons with different colors. The SPDC process couples the qubits in pairs (one of each color). The pairs can then be arbitrarily connected as long as the connections are between qubits with the same color. With only two photons, this can already enable the realization of various interesting graph states, such as the ones presented in fig.\ref{fig:S6}. Note that here, depending on the dimension of encoding, each vertex can represent either a qubit or a qudit.

Clearly, large-scale quantum computation cannot be realized with only two photons. Yet the flexibility of cluster state generation using only two photons hints at the rich diversity of graph states that could be realized using additional photons, measurements, and high-dimensional fusion gates.

\subsection*{Scaling of the number of required planes for single-qubit measurements}

When the number of qubits encoded per photon increases, the main challenge becomes implementing the linear optic circuits that set the measurement bases of the different qubits. As the number of optical modes $M=2^N$ grows exponentially with the number of qubits encoded per photon, one might be concerned that the depth of the optical circuit and the number of parameters defining the transformation would grow exponentially as well. Indeed, an MPLC with $O(M=2^N)$ planes is generally required to realize arbitrary $MXM$ unitary transformations.

However, for MBQC, each qubit is measured either in the Z basis or on the equatorial plane of the Bloch sphere in a measurement basis $B(\theta)=\left\{\frac{\ket{0}+e^{i\theta}\ket{1}}{\sqrt{2}},\frac{\ket{0}-e^{i\theta}\ket{1}}{\sqrt{2}}\right\}$\cite{raussendorf2003measurement}. The unitary transformation defining the measurement bases of all qubits is thus in a separable form $U^{(1)}_{2X2} \otimes U^{(2)}_{2X2} \otimes \cdots \otimes U^{(N)}_{2X2}$, where $U^{(j)}_{2X2}$ is the $2X2$ unitary transformation defining the measurement basis of the jth qubit. Therefore, the number of parameters defining the measurement bases of all qubits and the associated linear optic transformation scales only linearly with the number of qubits $N$, and as we will show, so is the depth of the linear optic circuit. In fact, this is the main motivation behind mapping the high-dimensionally encoded information in a photon onto several qubits or qudits in the first place, as it naturally keeps the complexity of the linear optic implementation manageable.

\begin{figure}[H]
\centering
\includegraphics[width=1\textwidth]{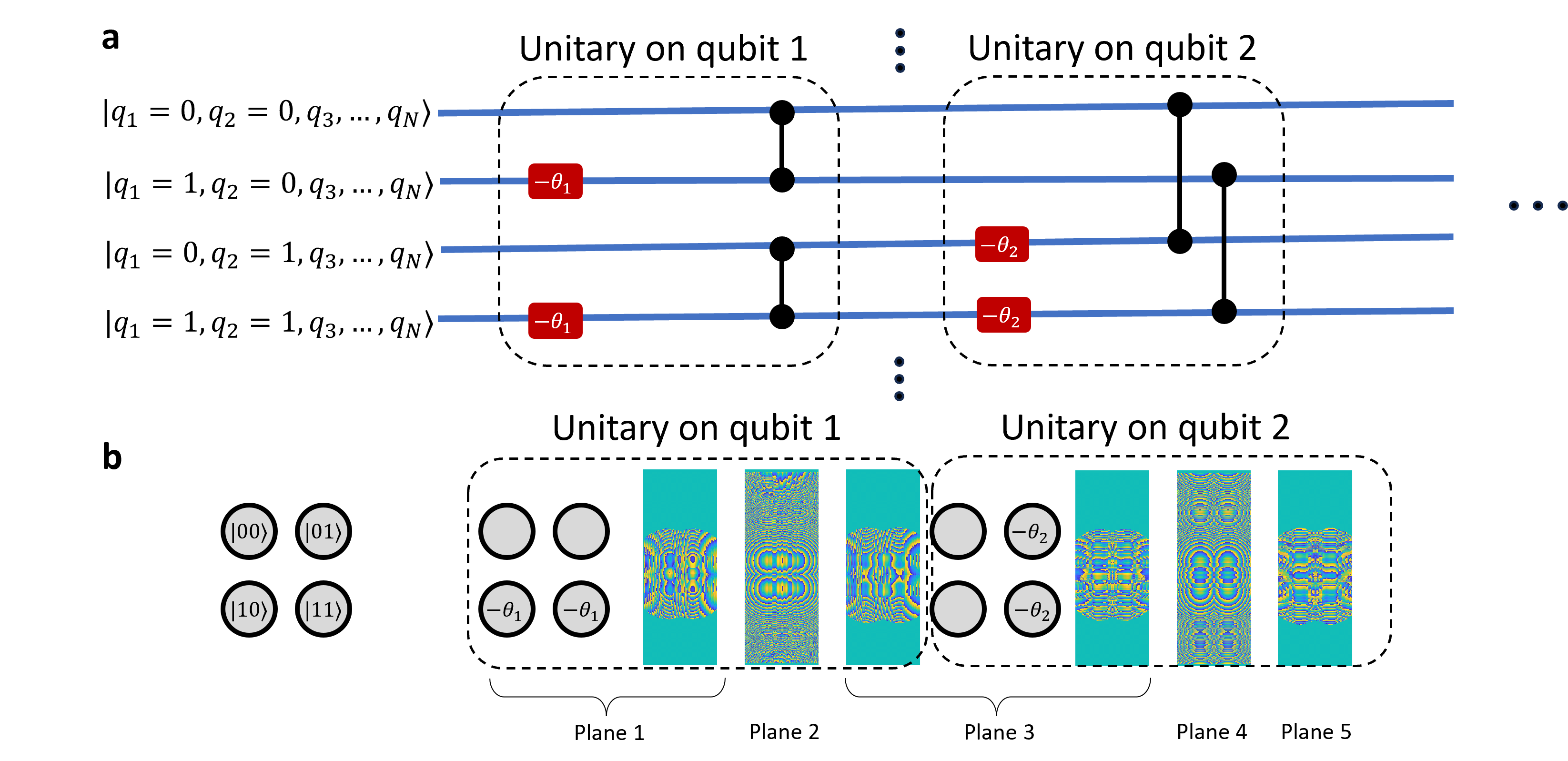}
\caption{\label{fig:S7} \textbf{Realizing measurements.} $N$ qubits are encoded in $2^N$ modes of a single photon. The measurement of each qubit $j$ in a basis $B(\theta_j)$ is implemented using appropriate phase shifters and beam splitters (see text). An example of measurement of the first two qubits $(q_1,q_2)$ out of $N$ qubits is presented in (a). The phase masks of its simulated MPLC realization are shown in (b).}
\end{figure}

To perform a measurement of a single qubit in the basis $B(\theta)$, one first needs to add a phase $-\theta$ to the modes associated with the $\ket{1}$ state of the measured qubit and then perform a beam-splitter operation between the modes encoding the state of the qubit. When multiple qubits are encoded per photon, $\frac{M}{2}$ phase shifts and beam-splitters are required, where $M$ is the number of optical modes per photon. Fortunately, these phase shifts and beam splitters can be implemented on all pairs of modes in parallel, making the depth of the linear optic circuit that realizes a measurement of a single qubit independent of the number of qubits encoded by each photon. Therefore, to realize a general measurement of $N$ qubits on the equatorial plane of the Bloch sphere, one could first perform the transformation for the first qubit, then the second, etc., making the depth of the entire circuit at most linear with the number of qubits $N$ encoded per photon, as shown in fig.\ref{fig:S7}a. This also holds true when taking intra-feedforward into account, as we discuss in the next section of the supplementary information.

To illustrate how this can be implemented via MPLC, we use the example of two qubits encoded in four modes of a single photon and simulate how to realize arbitrary measurements of both qubits on the equatorial plane of the Bloch sphere. In this case, to measure the first qubit in a basis $B(\theta_1)$, we add a phase shift of $-\theta_1$ to modes $\ket{10},\ket{11}$ and then perform beam splitter operations between the pairs of modes $\ket{00},\ket{10}$ and $\ket{01},\ket{11}$. Similarly, for the second qubit, we add a phase shift of $-\theta_2$ to mode $\ket{01},\ket{11}$ and then perform beam splitter operations between the pairs of modes $\ket{00},\ket{01}$ and $\ket{10},\ket{11}$. With the parameters of our MPLC, we find that the beam splitter operations can be implemented using only three planes with $99.8\%$ fidelity. The phase shifters are easily implemented by changing the phase of the SLM pixels at the first plane of the beam-splitter implementation. As the last plane used in the transformation of the first qubit and the first plane used in the transformation of the second qubit can be combined, the complete measurement can be realized with five planes in this case (see fig.\ref{fig:S7}b).

As the number of planes is expected to grow linearly with the number of qubits, the main challenge left in scaling up the number of qubits per photon would be the number of modes that should be manipulated in each plane. In our experiment, taking into account the size of the phase masks needed to manipulate the 50 spatial modes of the two photons, we were limited to 10 planes due to the size of our SLM. However, this can be resolved using additional SLMs or, preferably, using fixed phase masks or photonic integrated circuits\cite{tanomura2022scalable}, which could be scaled more easily. Indeed, MPLC is particularly promising for this task of manipulating a large number of modes using a few planes. For example, the multiplexing of over 1000 Hermite-Gaussian modes has been recently demonstrated using only 14 fixed phase masks\cite{fontaine2021hermite}. We envision that such demonstrations could be combined with programmable phase shifters to increase the number of manipulated modes and reduce the loss of the device. Note that for measurement bases that do not require feedforward (see next section), all the programmable phase shifts could be applied at the first plane, using for example, a single SLM, followed by a set of fixed phase masks. 

\subsubsection*{Loss scaling}

Given an effective transmission $t$ of each plane of the MPLC, the fact that the number of planes scales linearly with the number of qubits yields a total transmission through the MPLC that scales as $O(t^N)$ (rather than $O(t^{2^N})$ for a general unitary). This is comparable to the probability that $N$ photons that encode $N$ qubits will pass through a linear optic circuit with transmission $t$ per photon. However, as we show experimentally, the detection rate could still be significantly higher, as other probabilistic processes, such as photon generation, detection, and entanglement, now scale with the number of photons rather than qubits. 

Eventually, the effect of loss should be mitigated via quantum error correction. Our approach, where multiple qubits are encoded per photon, yields a unique error model in which qubits are lost (or not) in a correlated manner. The optimal implementation of quantum error correction remains an open question in this case, and we hope our work will motivate further research on fault-tolerant MBQC with high-dimensional encoding.

\subsection*{Realization of a CNOT gate}

\begin{figure}[H]
\centering
\includegraphics[width=0.9\textwidth]{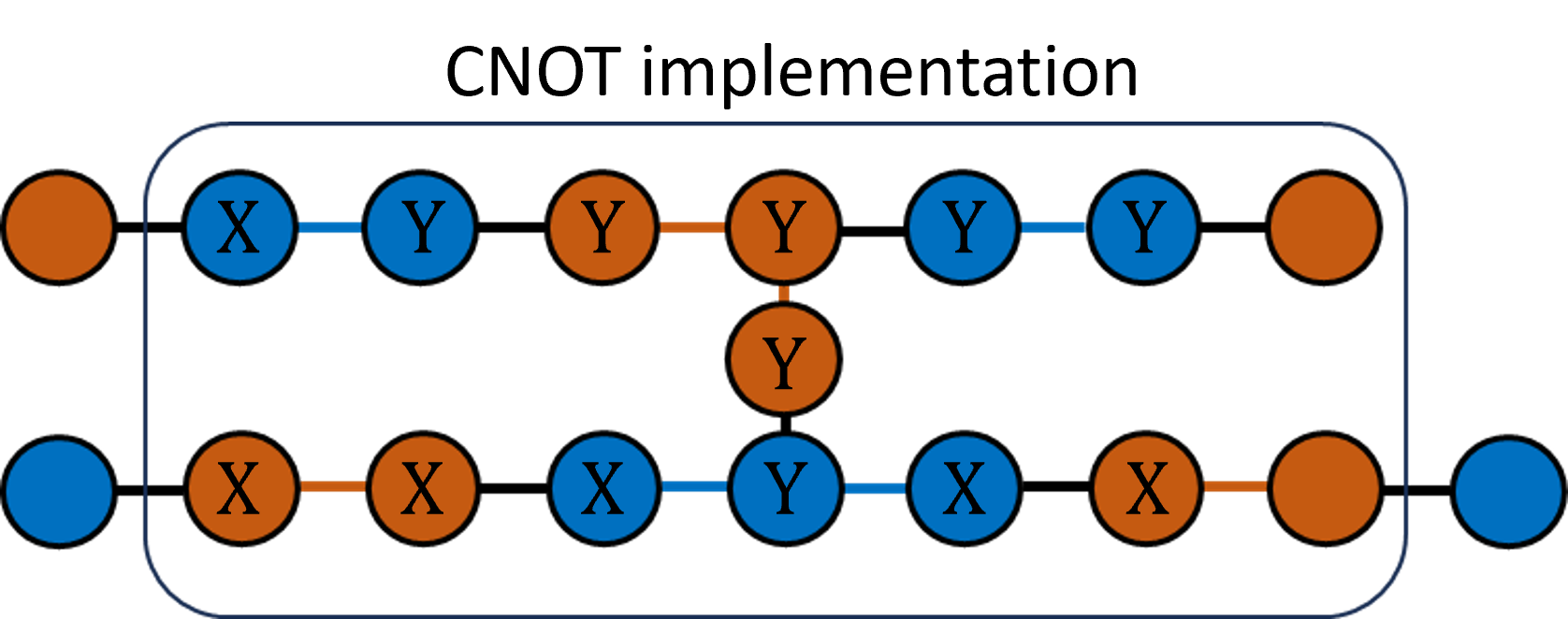}
\caption{\label{fig:S8} \textbf{Graph state for CNOT gate demonstration.} A CNOT gate in MBQC is realized by measuring qubits in the $X$ and $Y$ bases, according to the presented pattern. As in fig.\ref{fig:S6}, blue and orange vertices are associated with different photons.}
\end{figure}

In MBQC, a universal set of gates consisting of arbitrary single-qubit gates and a two-qubit CNOT gate is often considered\cite{raussendorf2003measurement}. As explained in the main text, in the experiment we focused on implementing single-qubit gates since non-Clifford gates require feedforward, making their realization different when multiple qubits are encoded on a single photon. In contrast, the two-qubit CNOT gate is a Clifford gate, requiring only Pauli measurements without adaptation of the measurement bases. Therefore, while requiring more qubits, the realization of the CNOT gate is conceptually simpler than non-Clifford gates and could be demonstrated, for example, using two photons encoding the graph state shown in fig.\ref{fig:S8}\cite{raussendorf2003measurement}.

\subsection*{Scaling instantaneous intra-feedforward to complete quantum circuits}

In the experiment, we demonstrate the realization of Clifford and non-Clifford single-qubit gates. For non-Clifford gates involving several qubits encoded onto the same photon, we have proposed and demonstrated the idea of intra-feedforward, where we use tailored linear optic circuits to mimic the effect of several feedforward steps in standard MBQC at a single time step. In this section, we discuss the generalization of this concept from single-qubit gates to complete quantum circuits. We first explore the applicability of intra-feedforward when all qubits in the computation are encoded onto a single photon, and then study the conditions for achieving deterministic quantum computation when intra- and inter-feedforward steps are combined.

\subsubsection*{Arbitrary feedforward between qubits encoded onto the same photon}

Before studying the general feedforward scenario involving multiple qubits and photons, it would be insightful to consider the extreme case of a complete quantum circuit where all qubits ($q_1,\cdots,q_N$) are encoded onto a single photon. Assuming the qubits $q_1,\cdots,q_N$ are ordered according to the order of their measurement, the measurement basis of a qubit $q_j$ might depend on the measurement results $\{m_i\}_{i<j}$ of previous qubits. As Pauli measurements (including Pauli $Z$ measurements) are not adapted in the computation, the general measurement basis of a qubit $q_j$ that depends on the measurement results of previous qubits would be on the equatorial plane of the Bloch sphere, having the form\cite{raussendorf2003measurement}

\begin{equation}
B(f_j(m_1,\cdots,m_{j-1})\theta_j) = \left\{\frac{\ket{0}+e^{if_j\theta_j}\ket{1}}{\sqrt{2}},\frac{\ket{0}-e^{if_j\theta_j}\ket{1}}{\sqrt{2}}\right\}
\end{equation}

where $f_j(m_1,\cdots,m_{j-1})=\pm 1$ determines the sign of the angle of measurement according to the previous measurement results and the structure of the feedforward circuit.

When all qubits are encoded onto a single photon, we have $M=2^N$ optical modes, each associated with a state $\ket{q_1,\cdots,q_N}$. To measure a given qubit $j$ at the right basis, we need to add a phase $-f_j(m_1,\cdots,m_{j-1})\theta_j$ to modes $\ket{q_1,\cdots,q_{j-1},1,q_{j+1},\cdots,q_N}$ where the jth qubit has a value of one, and then apply beam splitter operations between modes originally associated with $\ket{q_1,\cdots,q_{j-1},0,q_{j+1},\cdots,q_N}$ and $\ket{q_1,\cdots,q_{j-1},1,q_{j+1},\cdots,q_N}$. While the previous qubits are not actually 'measured' in the regular sense, their measurement value for each optical mode is known and is given according to $m_i=(-1)^{q_i}$ for a mode originally associated with qubit value $q_i$.

\begin{figure}[H]
\centering
\includegraphics[width=0.9\textwidth]{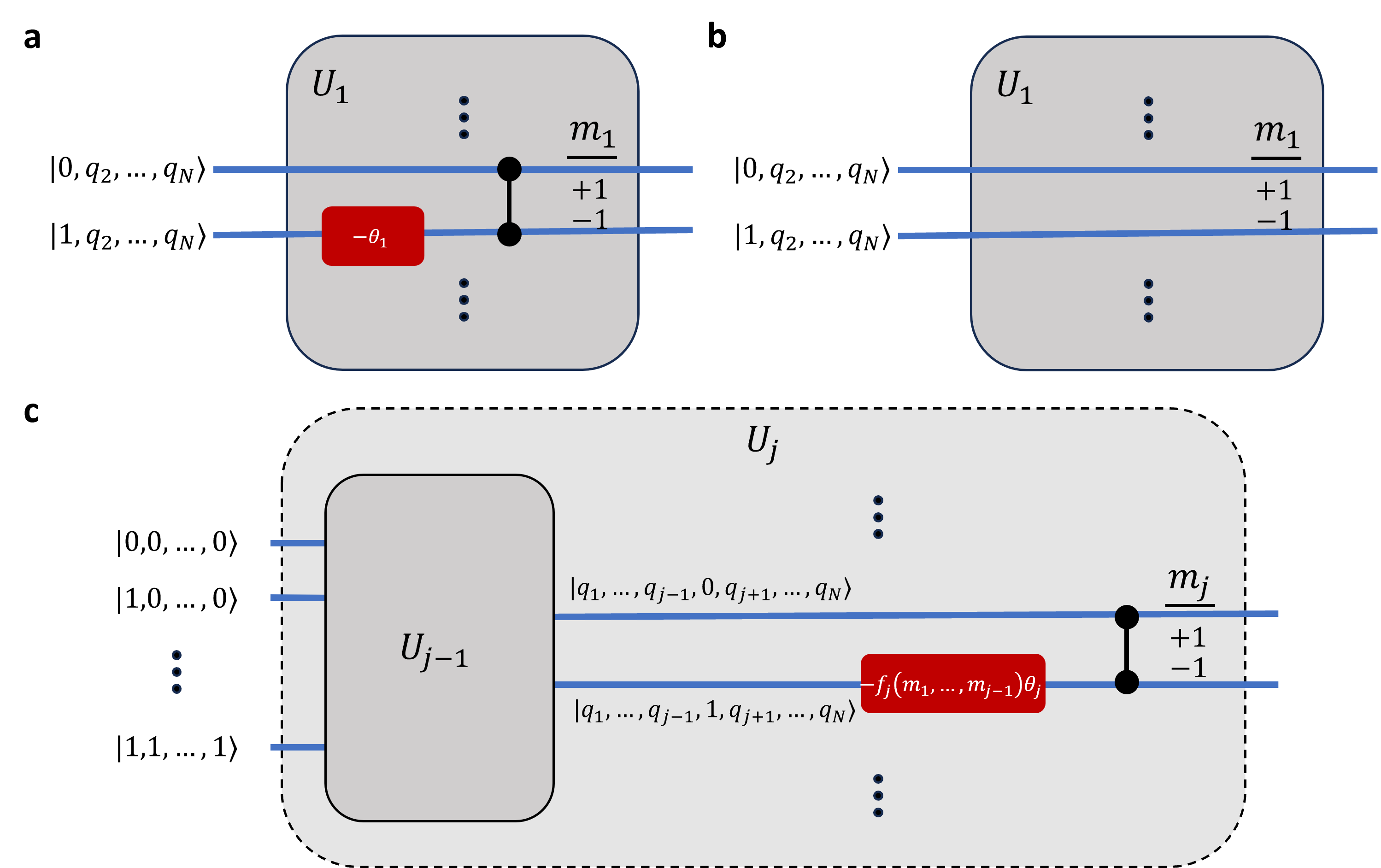}
\caption{\label{fig:S9} \textbf{Intra-feedforward for a complete quantum circuit.} The first qubit is measured either on the equatorial plane of the Bloch sphere (a) or in the $Z$ basis (b). In both cases, the measurement setting does not depend on the results of other qubits. Assuming the first $j-1$ qubits have been measured using a linear optic circuit $U_{j-1}$, the jth qubit can be measured in a basis dependent on their measurement results (c).}
\end{figure}

In fig.\ref{fig:S9}, we show by induction how a complete circuit can be constructed. The measurement of the first qubit is not dependent on previous measurements and can be easily performed in either the equatorial plane of the Bloch sphere (fig.\ref{fig:S9}a) or in the Pauli Z basis (fig.\ref{fig:S9}b). Assuming a unitary transformation $U_{j-1}$ can perform a general measurement of $j-1$ qubits encoded onto the same photon, including intra-feedforward, the measurement results of these qubits are known for every optical mode at the output of $U_{j-1}$. Therefore, the measurement of the jth qubit at the right basis can be straightforwardly performed as described above and shown in fig.\ref{fig:S9}c.

Therefore, by induction, this shows that any quantum circuit of $N$ qubit can be realized using intra-feedforward, with a linear optic circuit of depth $O(N)$. The only difference from what is described in the section 'Scaling of the number of required planes for single-qubit measurements' above is that the phase shifts cannot be generally pushed back towards the input of the transformation through the beam splitters.

\subsubsection*{Combining inter- and intra-feedforward for quantum computation}

When multiple photons, each encoding several qubits, are involved in the computation, inter- and intra-feedforward must be combined to achieve deterministic computation. As we have just shown, intra-feedforward between qubits encoded onto the same photon can always be performed. Therefore, the challenge is to combine the two approaches in a single computation. 

To understand under which conditions the combination of approaches can be successfully implemented, it would be useful to recall how measurements are temporally ordered in standard MBQC with inter-feedforward\cite{raussendorf2003measurement,raussendorf2001computational}. The set of qubits $q_i$ whose measurement basis directly depends on the measurement result of qubit $q_j$ are said to be in the forward cone of the qubit $q_j$, $fc(q_j)$. Such forward cones impose a strict partial order $\prec$ on the measurement of the qubits, such that if $q_i\in fc(q_j)$ then $q_j\prec q_i$, meaning that qubit $q_j$ is measured before qubit $q_i$. Therefore, the entire cluster state can be divided into disjoint sets of qubits $Q_t$, where the measurement bases of all qubits in the set $Q_t$ do not depend on the measurement results of qubits done in rounds $t\leq t'$. Thus, all qubits in set $Q_t$ can be measured simultaneously in round $t$. Similarly, $Q^{(t)}$ is defined as the set of qubits that have yet to be measured after round $t-1$ (where $Q^{(1)}$ includes all the qubits in the cluster). These sets of qubits are given by\cite{raussendorf2001computational}

\begin{equation}
Q_t = \left\{q\in Q^{(t)}| \neg \exists p \in Q^{(t)}: p \prec q \right\}
\end{equation}
\begin{equation}
Q^{(t+1)} = Q^{(t)}\setminus Q_t
\end{equation}

where the number of such sets $Q_t$ determines the duration of the computation.

In our case, each photon encoded multiple qubits. Therefore, the sets of simultaneously measured qubits $Q_t$ do not directly translate to sets of simultaneously measured photons. We thus wish to similarly define sets $P_t$ of simultaneously measured \textbf{photons} rather than qubits. If such sets can be identified, standard inter-feedforward between them, together with intra-feedforward between qubits encoded onto the same photon, would enable deterministic computation. 

For this, we define the forward cone of the nth photon $p_n\in \mathbf{P}$ as the set of all other photons that have at least one qubit that is in the forward cone of a qubit encoded onto the nth photon:

\begin{equation}
fc_{ph}(p_n) = \left\{p_m\in \mathbf{P}\setminus p_n| \exists (q_j \in p_n \ and\  q_i \in p_m): q_i\in fc(q_j) \right\}
\end{equation}

where $\mathbf{P}\setminus p_n$ is the set of all photons encoding the cluster state except the nth photon $p_n$. 

Given the notion of a forward cone of a photon, we can try and define a photonic strict partial order $\prec_{ph}$ on the measurements of the photons, where if $p_m\in fc_{ph}(p_n)$ then $p_n\prec_{ph} p_m$, meaning that the nth photon must be measured before the mth photon. However, for $\prec_{ph}$ to be a strict partial order, we must verify it is transitive (i.e. if $p_a \prec_{ph} p_b$ and $p_b \prec_{ph} p_c$ then $p_a \prec_{ph} p_c$) and anti-reflexive (i.e. $\neg \exists p\in\mathbf{P}$ such that $p\prec_{ph} p$). 

When each photon encodes one qubit, the transitivity and anti-reflexivity conditions are trivially met. If a qubit $q_a$ must be measured before qubit $q_b$ and qubit $q_b$ must be measured before qubit $q_c$, then of course qubit $q_a$ must be measured before qubit $q_c$, making $\prec$ transitive. In addition, the measurement basis of a qubit does not depend on its own measurement result, making $\prec$ anti-reflexive. However, when multiple qubits are encoded onto each photon, setting a legitimate strict partial order on the measurement of the photons becomes more subtle. Take for example a case where qubits $q_a$ and $q_c$ are encoded onto photon $p_n$ ($q_a,q_c \in p_n)$ and qubit $q_b$ is encoded in photon $p_m$ ($q_b\in p_m$). When $q_b\in fc(q_a)$ and $q_c\in fc(q_b)$, we have $p_m\in fc_{ph}(p_n)$ and $p_n\in fc_{ph}(p_m)$, yielding $p_n\prec_{ph} p_m$ and $p_m\prec_{ph} p_n$. From transitivity, this yields $p_n\prec_{ph} p_n$, which contradicts anti-reflexivity, preventing the establishment of strict partial order.

To make sure we can set a proper strict partial order on the measurement of the photons, we need to verify that all qubits encoded onto every pair of photons 'agree' on the order of measurement of these two photons. This can be formalized using the following condition, stating that if at least one qubit in photon $p_m$ must be measured after a qubit in a different photon $p_n$, then there cannot be another pair of qubits encoded in photons $p_m$ and $p_n$ that need to be measured in the opposite order:

\begin{equation}
\label{cond}
\begin{aligned}
\forall n \neq m \text {, if } \exists & \left(q_i \in p_n \text { and } q_j \in p_m\right) \text { such that } \\ q_i 
& \prec q_j \text {, then } \neg \exists\left(q_k \in p_n \text { and } q_l \in p_m\right) \text { such that } q_l \prec q_k
\end{aligned}
\end{equation}

When this condition is satisfied, transitivity and anti-reflexivity are inherited from $\prec$, and $\prec_{ph}$ defines a legitimate strict partial order on the photonic measurements, yielding a set of simultaneously measured photons $P_t$ in step $t$:

\begin{equation}
P_t = \left\{p\in P^{(t)}| \neg \exists p' \in P^{(t)}: p' \prec_{ph} p \right\}
\end{equation}
\begin{equation}
P^{(t+1)} = P^{(t)}\setminus P_t
\end{equation}

We note that the condition in eq.\ref{cond} can always be satisfied, for example, by encoding all qubits onto a single photon, all qubits within each set $Q_t$ onto a single photon, or each qubit onto a different photon. However, finding the ideal allocation of qubits onto photons that reduces the number of measurement steps while still satisfying condition \ref{cond} is not trivial and depends on the limitation of a given physical implementation. We thus leave this as an open question for future study and optimization.

\end{document}